\documentstyle[psfig]{mn}

\newif\ifAMStwofonts
\AMStwofontstrue

\newcommand{\be}{\begin{equation}}
\newcommand{\ee}{\end{equation}}
\newcommand{\ba}{\begin{eqnarray}}
\newcommand{\ea}{\end{eqnarray}}
\newcommand{\brr}{\begin{array}}
\newcommand{\err}{\end{array}}
\newcommand{\bc}{\begin{center}}
\newcommand{\ec}{\end{center}}

\newcommand{\lum}{\,{\rm erg\,s^{-1}}}
\newcommand{\lt}{$L_X$--$T$$~$}

\newcommand{\mincir}{\raise
  -2.truept\hbox{\rlap{\hbox{$\sim$}}\raise5.truept \hbox{$<$}\ }}
\newcommand{\magcir}{\raise
  -2.truept\hbox{\rlap{\hbox{$\sim$}}\raise5.truept \hbox{$>$}\ }}
\newcommand{\siml}{\raise
  -2.truept\hbox{\rlap{\hbox{$\sim$}}\raise5.truept \hbox{$<$}\ }}
\newcommand{\simg}{\raise
  -2.truept\hbox{\rlap{\hbox{$\sim$}}\raise5.truept \hbox{$>$}\ }}

\ifoldfss
\ifCUPmtlplainloaded \else
\NewTextAlphabet{textbfit} {cmbxti10} {}
\NewTextAlphabet{textbfss} {cmssbx10} {}
\NewMathAlphabet{mathbfit} {cmbxti10} {} 
\NewMathAlphabet{mathbfss} {cmssbx10} {} 
\fi
\ifAMStwofonts
\ifCUPmtlplainloaded \else
\NewSymbolFont{upmath} {eurm10}
\NewSymbolFont{AMSa} {msam10}
\NewMathSymbol{\upi}     {0}{upmath}{19}
\NewMathSymbol{\umu}     {0}{upmath}{16}
\NewMathSymbol{\upartial}{0}{upmath}{40}
\NewMathSymbol{\leqslant}{3}{AMSa}{36}
\NewMathSymbol{\geqslant}{3}{AMSa}{3E}

 \let\le=\leqslant
 \let\ge=\geqslant
\fi
\fi
\fi 

\ifnfssone
\newmathalphabet{\mathit}
\addtoversion{normal}{\mathit}{cmr}{m}{it}
\addtoversion{bold}{\mathit}{cmr}{bx}{it}
\newmathalphabet{\mathbfit} 
\addtoversion{normal}{\mathbfit}{cmr}{bx}{it}
\addtoversion{bold}{\mathbfit}{cmr}{bx}{it}
\newmathalphabet{\mathbfss} 
\addtoversion{normal}{\mathbfss}{cmss}{bx}{n}
\addtoversion{bold}{\mathbfss}{cmss}{bx}{n}
\ifAMStwofonts
\ifCUPmtlplainloaded \else
\UseAMStwoboldmath
\makeatletter
\new@mathgroup\upmath@group
\define@mathgroup\mv@normal\upmath@group{eur}{m}{n}
\define@mathgroup\mv@bold\upmath@group{eur}{b}{n}
\edef\UPM{\hexnumber\upmath@group}
\new@mathgroup\amsa@group
\define@mathgroup\mv@normal\amsa@group{msa}{m}{n}
\define@mathgroup\mv@bold\amsa@group{msa}{m}{n}
\edef\AMSa{\hexnumber\amsa@group}
\makeatother
\mathchardef\upi="0\UPM19
\mathchardef\umu="0\UPM16
\mathchardef\upartial="0\UPM40
\mathchardef\leqslant="3\AMSa36
\mathchardef\geqslant="3\AMSa3E

 \let\le=\leqslant
 \let\ge=\geqslant
\fi
\fi
\fi 

\ifnfsstwo
\DeclareMathAlphabet{\mathbfit}{OT1}{cmr}{bx}{it}
\SetMathAlphabet\mathbfit{bold}{OT1}{cmr}{bx}{it}
\DeclareMathAlphabet{\mathbfss}{OT1}{cmss}{bx}{n}
\SetMathAlphabet\mathbfss{bold}{OT1}{cmss}{bx}{n}
\ifAMStwofonts
\ifCUPmtlplainloaded \else
\DeclareSymbolFont{UPM}{U}{eur}{m}{n}
\SetSymbolFont{UPM}{bold}{U}{eur}{b}{n}
\DeclareSymbolFont{AMSa}{U}{msa}{m}{n}
\DeclareMathSymbol{\upi}{0}{UPM}{"19}
\DeclareMathSymbol{\umu}{0}{UPM}{"16}
\DeclareMathSymbol{\upartial}{0}{UPM}{"40}
\DeclareMathSymbol{\leqslant}{3}{AMSa}{"36}
\DeclareMathSymbol{\geqslant}{3}{AMSa}{"3E}

 \let\le=\leqslant
 \let\ge=\geqslant
\fi
\fi
\fi 

\ifCUPmtlplainloaded \else
\ifAMStwofonts \else 
\def\upi{\pi}
\def\umu{\mu}
\def\upartial{\partial}
\fi
\fi

\title[] {The effect of non--gravitational gas heating in groups and
  clusters of galaxies} 
\author[Non--gravitational gas heating]{
S. Borgani$^1$, 
F. Governato$^{2,7}$, 
J. Wadsley$^3$,
N. Menci$^4$, 
P. Tozzi$^5$, \\~\\
\LARGE{
T. Quinn$^2$, 
J. Stadel$^6$,
G. Lake$^2$
}
\\ ~\\ 
$^1$ INFN, Sezione di Trieste, c/o Dipartimento di Astronomia
dell'Universit\`a, via Tiepolo 11, I-34131 Trieste, Italy
(borgani@ts.astro.it)\\ 
$^2$ Astronomy Department, University of Washington, Seattle WA 98195,
USA (fabio, trq, lake @hermes.astro.washington.edu)\\ 
$^3$ Department of Physics and Astronomy, McMaster University,
Hamilton, Ontario, L88 4M1, Canada (wadsley@physics.mcmaster.ca)\\
$^4$ INAF, Osservatorio Astronomico di Roma, via dell'Osservatorio, I-00040
Monteporzio, Italy (menci@coma.mporzio.astro.it) \\ 
$^5$ INAF, Osservatorio Astronomico di Trieste, via Tiepolo 11, I-34131
Trieste, Italy (tozzi@ts.astro.it)\\
$^6$ University of Victoria, Department of Physics and Astronomy, 3800
Finnerty Road, Elliot Building, Victoria, BC V8W 3PG Canada
\\$~~~$(stadel@phys.uvic.ca)\\
$^7$ INAF, Osservatorio Astronomico di Brera, via
Brera 28, I-20131, Milano, Italy \\ 
}
\date{}
\pubyear{2002}

\begin{document}
\label{firstpage}
\maketitle

\begin{abstract}
  We present a detailed study of a set of gas-dynamical simulations of
    galaxy groups and clusters in a flat, $\Lambda$CDM model with
    $\Omega_m=0.3$, aimed at exploring the effect of
    non--gravitational heating on the observable properties of the
    intracluster medium (ICM). We use {\tt{GASOLINE}}, a version of
    the code {\tt{PKDGRAV}} that includes an SPH description of
    hydrodynamics to simulate the formation of four cosmic halos with
    virial temperatures in the range $0.5\mincir T\mincir 8$ keV.
    These simulations resolve the structure and properties of the
    intra--cluster medium (ICM) down to a small fraction of the virial
    radius, $R_{\rm vir}$. At our resolution $X$--ray luminosities,
    ($L_X$), of runs with gravitational heating only are in good
    agreement,  over almost two orders of magnitude in
    mass, with analytical predictions, that assume a universal
    profile for CDM halos.

  For each simulated structure, non--gravitational heating of the ICM
  is implemented in two different ways: (1) by imposing a minimum
  entropy floor, $S_{fl}$, at a given redshift, that we take in the
  range 1$\le z\le$5; (2) by gradually heating gas within collapsed
  regions, proportionally to the supernova rate expected from
  semi--analytical modeling of galaxy formation in halos having mass
  equal to that of the simulated systems.

  Our main results are the following. {\em (a)} An extra heating
  energy $E_h\magcir 1$ keV per gas particle within $R_{vir}$ at $z=0$
  is required to reproduce the observed $L_X$--$T$ relation,
  independent of whether it is provided in an impulsive way to create
  an entropy floor $S_{fl}=50$--100 keV cm$^2$, or is modulated in
  redshift according to the star formation rate; our SN feedback
  recipe provides at most $E_h\simeq 1/3$ keV/part and, therefore, its
  effect on the \lt relation is too small to account for the observed
  \lt relation. {\em (b)} The required heating implies, in small
  groups with $T\sim 0.5$ keV, a baryon fraction as low as $\mincir
  40\%$ of the cosmic value at $R_{\rm vir}/2$; this fraction
  increases to about 80\% for a $T\simeq 3$ keV cluster. {\em (c)}
  Temperature profiles are almost scale free across the whole explored
  mass range, with T decreasing by a factor of three at the virial
  radius.  {\em (d)} The mass--temperature relation is almost
  unaffected by non--gravitational heating and follows quite closely
  the $M\propto T^{3/2}$ scaling; however, when compared with data on
  the $M_{500}$--$T_{ew}$ relation, it has a $\sim 40\%$ higher
  normalization. This discrepancy is independent of the heating scheme
  adopted. The inclusion of cooling in a run of a small group steepens
  the central profile of the potential well while removing gas from
  the diffuse phase. This has the effects of increasing $T_{ew}$ by
  $\sim 30\%$, possibly reconciling the simulated and the observed
  $M_{500}$--$T_{ew}$ relations, and of decreasing $L_X$ by $\sim
  40\%$. However, in spite of the inclusion of SN feedback energy,
  almost 40\% of the gas drops out from the hot diffuse phase, in
  excess of current observational estimates of the amount of cold
  baryons in galaxy systems.

  Likely, only a combination of different heating sources (SNe and
  AGNs) and cooling will be able to reproduce both the $L_X$--$T_{ew}$
  and $M_{500}$--$T_{ew}$ relations, as observed in groups and
  clusters, while balancing the cooling runaway.

\end{abstract}


\section{Introduction}

The $X$--ray emission from clusters of galaxies offers a unique means
for studying the physics of cosmic baryons and their connection with
the processes of galaxy formation and evolution. 
The first attempt to model the ICM in the framework of the
hierarchical clustering scenario assumed its thermodynamical
properties to be entirely determined by gravitational processes, such as
adiabatic compression during collapse and shock heating by supersonic
gas accretion (Kaiser 1986). Since gravity does not have
characteristic scales, for an Einstein--de-Sitter cosmology this model
predicts rich massive clusters and poor groups to appear as scaled
versions of each other.  Under the assumptions of emissivity dominated
by free--free bremsstrahlung and of hydrostatic equilibrium of the
gas, this model predicts $L_X\propto T^2(1+z)^{3/2}$ for the shape and
evolution of the relation between $X$--ray luminosity and gas
temperature. Furthermore, if we define the gas entropy as
$S=T/n_e^{2/3}$ ($n_e$: electron number density; e.g. Eke et
al. 1998), then the self--similar ICM has $S\propto T(1+z)^{-2}$.

This simple model was quickly recognized to fail at accounting for
several observational facts: (a) the $L_X$--$T$ relation for nearby
clusters is steeper than predicted, with $L_X\propto T^{\sim 3}$ for
$T\magcir 2$ keV clusters (e.g., David et al. 1993, White, Jones \&
Forman 1997, Allen \& Fabian 1998, Markevitch 1998, Arnaud \& Evrard
1999), with a possible further steepening at the group scale,
$T\mincir 1$ keV (Ponman et al. 1996, Helsdon \& Ponman 2000); (b) no
evidence of evolution for its amplitude has been detected out to
$z\magcir 1$ (e.g., Mushotzky \& Scharf 1997, Donahue et al. 1999,
Fairley et al. 2000, Della Ceca et al. 2000, Borgani et al. 2001b,
Holden et al. 2002); (c) the gas density profile in central regions
of cooler groups is relatively softer than in clusters and,
correspondingly, the entropy is higher than predicted by self--similar
scaling (e.g., Ponman, Cannon \& Navarro 1999, Lloyd--Davis et
al. 2000, Finoguenov et al. 2002); (d) colder clusters contain a
relatively smaller amount of gas than very hot ones (e.g. Neumann \&
Arnaud 2001).

A common interpretation for such observational facts requires
non--gravitational energy input, which should have taken place during
the past history of the ICM. This extra heating would place the gas on
a higher adiabat and sustain the gas during the gravitational collapse
of the cluster dark matter (DM) halo, preventing it from reaching high
density in central regions and suppressing the $X$--ray emission. For
a fixed specific amount of heating energy per gas particle, the effect
is larger for the smaller systems, while being negligible for more
massive, hotter systems. Kaiser (1991) suggested for the first time
that gas pre--heating and subsequent adiabatic collapse turns into a
differential steepening of the $L_X$--$T$ relation at the cluster
scales. Evrard \& Henry (1991), Navarro, Frenk \& White (1995) and
Bower (1997) assumed the gas in the cluster core to have a minimum
entropy level, which was established at some pre--heating epoch.

Cavaliere, Menci \& Tozzi (1998, 1999) were able to predict the
correct $L_X$--$T$ slope, both at the cluster and at the group scale,
by assuming a pre--heated gas at a temperature $T\sim 0.5$ keV, which
is shocked with different strengths when falling into groups and
clusters. A pre--heating producing an isentropic gas distribution has
been assumed by Balogh, Babul \& Patton (1999) and Tozzi \& Norman
(2001, TN01 hereafter). In particular, TN01 worked out a series of
predictions of observables properties of the ICM with isentropic
pre--heating, including the amount of energy feedback required to
produce the correct \lt relation and entropy threshold at the group
scale. Brighenti \& Mathews (2001) discussed the effect of heating the
gas either when it has still to collapse within clusters and groups
(external heating) or when it is already accreted (internal heating).
They concluded that internal heating is more efficient at accounting
at the same time for both the slope of the \lt relation and the excess
entropy in poor clusters.

While most of such analysis agree that an extra heating energy of
$\sim 0.5$--1 keV per gas particle is required, yet no general
consensus has been reached about its astrophysical origin. Supernovae
(SNe) have been advocated by several authors as a possible source for
pre--heating (e.g., Wu, Fabian \& Nulsen 1998) and metal enriching
(e.g. Loewenstein \& Mushotzky 1996, Renzini 1997, Finoguenov \&
Ponman 1999, Pipino et al. 2002) the ICM. Although some analyses
showed that SNe can actually provide an adequate energy budget (Menci
\& Cavaliere 2000, Lloyd--Davis et al. 2000), other authors claimed
that this requires a very high, possibly unrealistic, efficiency for
the thermalization of the released energy (Wu, Fabian \&
Nulsen 2000, Kravtsov \& Yepes 2000, Bower et al. 2001).

In this case, alternative sources for ICM heating are required, like
SN from a primordial star population, the so--called Pop III stars
(e.g., Loewenstein 2001), and nuclear galactic activity associated with
QSOs (e.g., Valageas \& Silk 1999, Mc Namara et al. 2000, Yamada \&
Fujita 2001, Nath \& Roychowdhury 2002). As for Pop III stars, one
expects them not to significantly heat the diffuse medium, in order
not to destroy Ly--$\alpha$ absorbers at lower redshift, $z\mincir 3$,
and not to overpollute the IGM with metals.

To date, only few attempts have been pursued to address in detail the
effects of non gravitational heating on $X$--ray observable quantities
with numerical hydrodynamical simulations (Navarro, Frenk \& White
1995, Lewis et al. 2000, Borgani et al. 2001a, Valdarnini
2002). Bialek, Evrard \& Mohr (2001) have realized a large set of
moderate resolution simulations of clusters and groups, setting
different entropy floors at very high redshift. They found that
agreement with observational constraints on $X$--ray cluster scaling
properties requires an entropy floor of about 100 keV cm$^2$, which,
at their heating redshift $z_h\simeq 21$, correspond to an extra
energy of about 0.2 keV/part.

As an alternative to extra--heating, radiative cooling has been also
suggested as a mechanism to break ICM self--similarity. In this case,
$X$--ray luminosity is suppressed as a consequence of the reduced
amount of gas in the hot diffuse phase (Pearce et al. 2000, Muanwong
et al. 2001). In addition, low--entropy gas in central cluster regions
is selectively removed from the hot phase as a consequence of its
short cooling time, thus leaving only high entropy gas (Voit \& Bryan
2001). However, cooling in itself is known to convert into a cold
stellar phase a large fraction of gas, $> 30\%$ (e.g., Lewis et
al. 2000, Dav\'e et al. 2001), much larger than the $\mincir 10\%$
baryon fraction locked into stars indicated by observations. This
underlines the need for extra heating to prevent overcooling of the
ICM (e.g.  Suginohara \& Ostriker 1998, Prunet \& Blanchard 1999,
Balogh et al.  2001).

The main aim of this paper is to study in detail the effect of
non--gravitational energy injection into the ICM and how much of this
energy is needed to correctly reproduce different $X$--ray observable
properties of groups and clusters of galaxies. In a previous paper
(Borgani, Governato, Wadsley et al. 2001a, Paper I hereafter) we
concentrated on the effect of non--gravitational heating on the
entropy pattern of the ICM. As a main conclusion, we showed that the
entropy excess in the central regions of poor clusters (Ponman et
al. 1999) requires an extra heating of about 1 keV/particle. This
paper presents an extended analysis of a larger set of simulations,
which include several schemes of ICM extra heating.  We will also show
results from a simulation of a poor group, which includes gas cooling
and a star formation algorithm to remove dense cold gas particles from
the SPH computation. This simulation will be used to discuss the r\^ole
that cooling plays in determining the ICM thermodynamics.

The plan of the paper is as follows. In Section 2 we give a short
description of the Tree+SPH code Gasoline and provide details about
the simulations. Section 3 describes the different recipes for
non--gravitational heating that we used. Section 4 is devoted to the
presentation of the results. In Section 5 we discuss the effect of
cooling, while a summary of our main results and our conclusions are
presented in Section 6.

\section{Numerical Method}

\subsection{The Code}
We use Gasoline, a new N-body/Smoothed Particle Hydrodynamics code.
Gasoline performs gravity and hydrodynamics operations using a binary
tree-structure with periodic boundary conditions for cosmology.  It
uses concurrent multiple timesteps for increased throughput and runs
in parallel on a wide variety of architectures using, e.g., pthreads,
shared memory or the Message Passing Interface (MPI).  The code
handles gravity, pressure gradients, hydrodynamic shocks, radiative
cooling, photoionizing UV background and was extended for this work to
incorporate external heating sources.  For a full description of
Gasoline, including extended tests of the code, see Wadsley, Stadel \&
Quinn (2002, preparation). For a description of the gravity code on
which Gasoline is based, see Stadel (2001). The external heating SN
described in Section 3 is treated within the code as an additional
component added locally to the standard SPH compressive heating
(``PdV'') term.

\begin{figure*}
\centering
\psfig{file=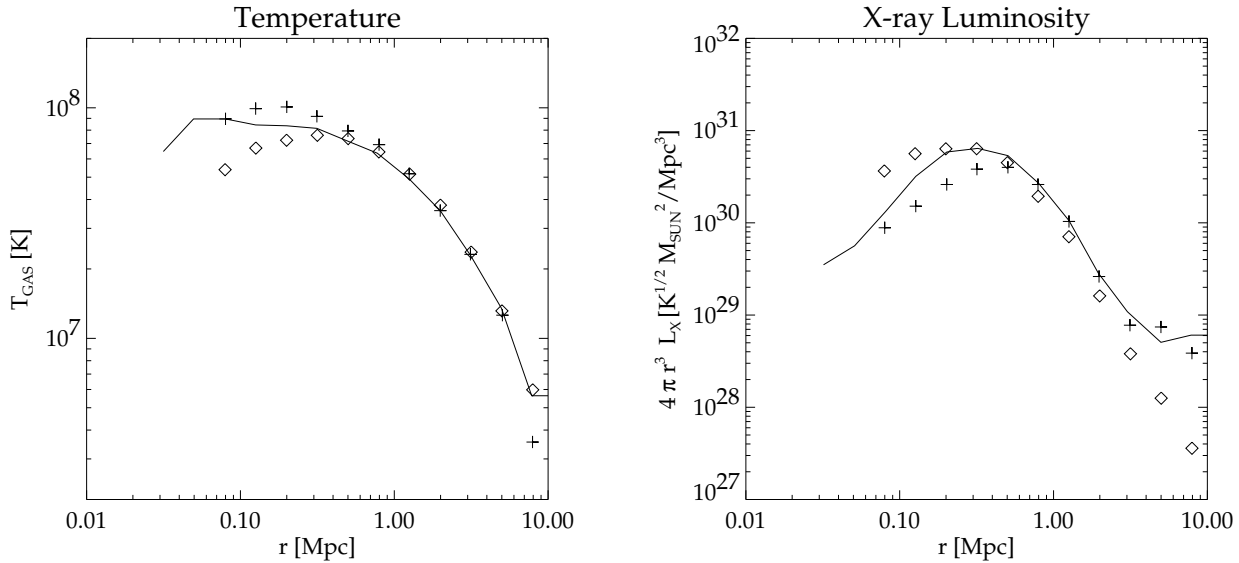,width=17.5truecm} 
\caption{``Cluster comparison'' gas profiles.  The Gasoline results at
$64^3$ (diamonds) for the temperature (left) and $X$--ray luminosity
(right) profiles are in excellent agreement with the averaged results
from different codes (solid line) and HYDRA (plus symbols) profiles
from the cluster comparison paper (Frenk at al. 2000).}
\label{cluscomp} 
\end{figure*}

As a demonstration of the effectiveness of Gasoline for cluster
simulations, we evolve the initial conditions from the ``Santa Barbara
Cluster Comparison'' (Frenk et al. 2000).  We resolve the resulting
luminous $X$--ray cluster with $\sim 15000$ gas particles within the
virial radius ($2.7$ Mpc) at the current epoch with $64^3$ gas and
$64^3$ dark matter particles in the whole box.  This resolution was
used by the majority of the cluster comparison SPH codes including the
widely used HYDRA code (Couchman 1991).  The test evolved the gas
adiabatically without additional heat sources or sinks.  The gas
profiles for Gasoline with curves for HYDRA and the average result
from the cluster comparison are shown in Figure~\ref{cluscomp}.  The
innermost radial bin plotted in each case is indicative of the
resolution.  The variation seen is similar to that found between
cluster comparison code results, even among different SPH codes, due
to differing implementations and the interpretation of the initial
condition data which were given to the coders at $256^3$ resolution.
The central temperature in particular is sensitive to shock treatment
and, most of all, minor variations in the timing of small subclump
mergers (for an extended discussion see Wadsley et al.  2002, in
preparation).  Except for these small variations, the Gasoline results
are in excellent agreement with the other codes.

\begin{figure*} 
\centerline{ \hbox{ \psfig{file=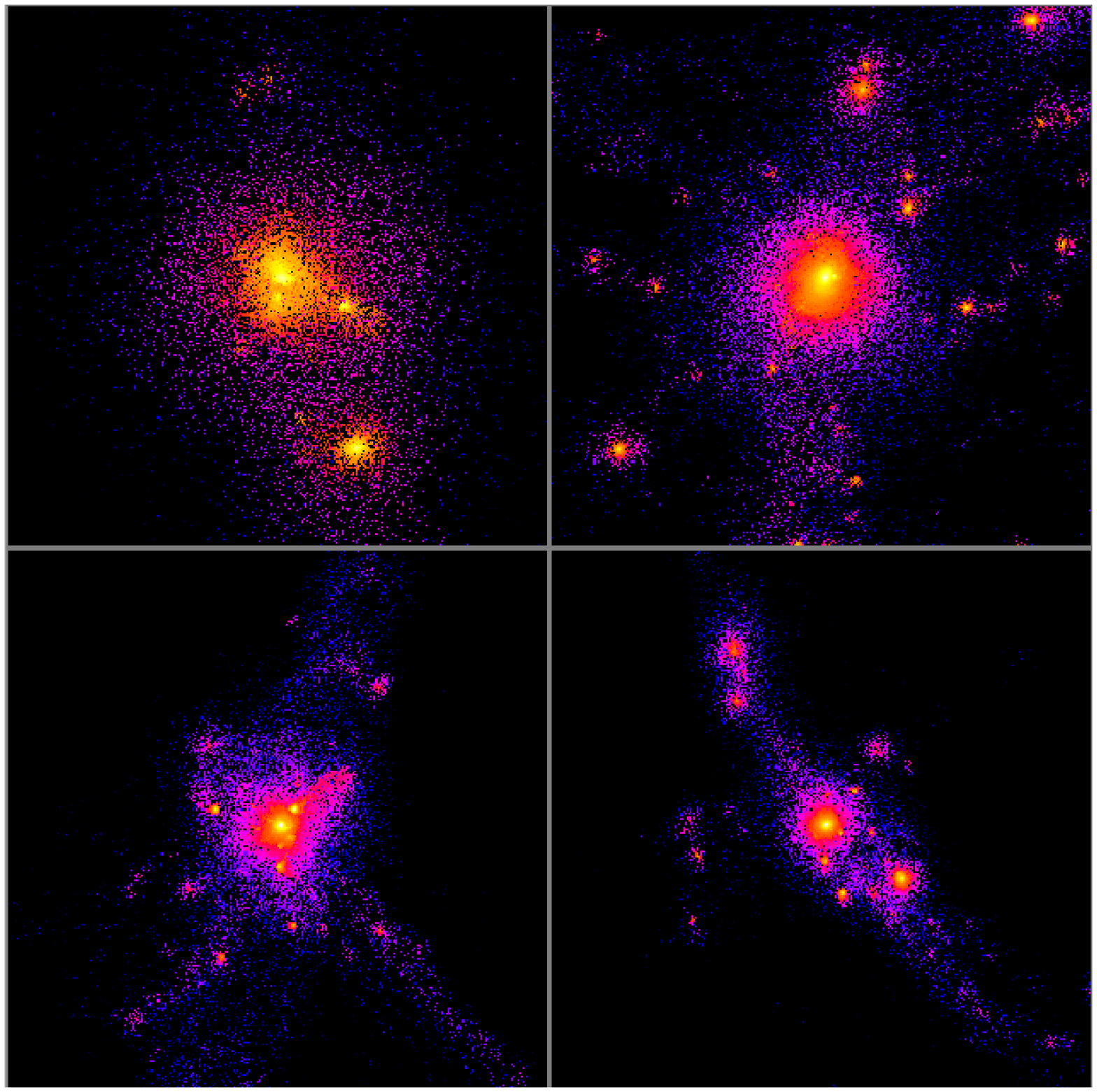,width=15.5truecm} }}
\caption{The map of the gas density for the HR version of the four
simulated structures. The Coma, Virgo, Fornax and Hickson runs
correspond to the upper left, upper right, lower left and lower right
panels, respectively. Each box correspond to a physical size of 10
Mpc.} 
\label{fi:gasmaps} 
\end{figure*}

\begin{table*}
\centering
\caption{Characteristics of the simulated  halos and numerical
parameters of the simulations. Column 2: simulation version (HR: high
resolution; LR: low resolution); Column 3: total mass within the
virial radius at z=0 ($10^{13}M_\odot$); Column 4: virial radius (Mpc);
Column 5: mass of gas particles ($10^8M_\odot$); Column 6: mass of DM
particles within the re-simulated region ($10^8M_\odot$); Column 7 and
8: number of gas and DM particles within the virial radius; Column 9:
 softening parameter (kpc; runs marked with $^*$
have been run also with a twice as large softening); Column 10:
starting redshift.}
\begin{tabular}{lccccccccc}
Run & & $M_{vir}$ & $R_{vir}$ & $m_{gas}$ & $m_{DM}$ & $N_{gas}$ & $N_{DM}$ & $\epsilon$ & $z_{in}$  \\
\\
Coma    &    & 133.6& 2.86 & 17.7 & 119.2& 0.99e5 & 0.90e5 & 7.5 & 49 \\
Virgo   & HR & 30.4 & 1.75 & 2.21 & 14.9 & 1.70e5 & 1.79e5 & 7.5 & 69 \\ 
        & LR &      &      & 17.7 & 119.2& 0.21e5 & 0.22e5 & 15  & 49 \\
Fornax  & HR & 5.91 & 1.01 & 0.65 & 4.41 & 1.06e5 & 1.18e5 &2.5$^*$& 89 \\ 
        & LR &      &      & 5.20 & 35.3 & 1.33e4 & 1.48e4 & 7.5 & 69 \\
Hickson & HR & 2.49 & 0.76 & 0.65 & 4.41 & 4.93e4 & 4.86e4 &2.5$^*$& 89 \\
        & LR &      &      & 5.20 & 35.3 & 0.62e4 & 0.61e4 & 7.5 & 69 \\

\end{tabular}
\label{t:simul}
\end{table*}

\subsection{The Simulations}
We simulated at high resolution four halos taken from a DM--only
cosmological simulation within a 100 Mpc box of a $\Lambda$CDM model
with $\Omega_m = 0.3$, $\Omega_\Lambda =0.7$, $\sigma_8$=1 and Hubble
constant $H_0=70$ km s$^{-1}$ Mpc$^{-1}$. The baryon density parameter
was chosen to be $\Omega_b=0.019\,h^{-2}$, as inferred from the
deuterium abundance detected in QSO absorption lines (e.g., Burles \&
Tytler 1998), which corresponds to a baryon fraction $f_{\rm
bar}\simeq 0.13$.

From the simulated box we selected four halos with masses
corresponding to virial temperatures in the range $0.5\mincir T\mincir
8$ keV. With this choice, we are sure to encompass the temperature
interval where the $L_X$--$T$ relation is observed to steepen
substantially and where the effect of energy injection should be more
important. The sizes of the four simulated structures are of the same
order as those of the Coma cluster (e.g. Geller, Diaferio \& Kurtz
1999), the Virgo cluster (Bingelli, Tammann \& Sandage 1987), the
Fornax group (e.g. Drinkwater et al. 2000) and of a smaller
Hickson--like group (e.g. Pildis, Bregman \& Evrard 1995). In the
following we indicate with these names the corresponding simulated
structures.

Using the so--called renormalization technique (Katz \& White 1983),
the simulation within the periodic box is centered around each
re--simulated halo, the mass distribution and the perturbation
spectrum are resampled at higher resolution within the Lagrangian
region that contains the halo of interest. The high--resolution region
extends typically for a few comoving Mpc around the halo. The outer
regions are resampled at decreasing resolution.  As the four halos
span almost two decades in mass, the numerical resolutions differ
accordingly, so as to keep comparable the number of particles within
the virial radius and the force resolution in units of the virial
radius. Typically 0.5--1.5 million particles are used for each run, with
about half of them to simulate the high resolution central regions. A
summary of the characteristics of the simulations are reported in
Table \ref{t:simul}. Note that halo particle number refers to the
number of particles within the virial radius, and not just in the high
resolution region, which typically contains a few times more
particles.  In order to investigate the effect of resolution, the
Virgo, Fornax and Hickson runs have been realized with different mass
and force resolution. The high--resolution version of each run has a
force softening which is at most 1\% of the virial radius, while the
mass resolution is chosen so that at least $5\times 10^4$ gas
particles fall within that radius by $z=0$. For all the runs we used
$\eta=0.2$ and a force accuracy parameter of the tree algorithm of
$\theta =0.7$ ($\theta =0.5$ for $z>2$), 
as suggested by Moore et al. (1998) and later verified
by Power et al. (2002), for the parameters regulating time
accuracy of the integration and the accuracy of the tree algorithm.

We show in Figure \ref{fi:gasmaps} the gas density maps for the
high--resolution regions of the simulated structures at $z=0$.  The
quite high resolution achieved allows us to resolve numerous
substructures. Some of them retain part of their individual gas
content, which survives for a few crossing times before being
stripped, as well as discontinuities associated with bow shocks (see
(see Paper I for a discussion of these features as seen from the
entropy maps of simulations). It is tempting to associate such
structures to the small--scale features revealed by Chandra
observations, which in some cases are associated with bow shocks or
cold fronts from merging structures (e.g., Markevitch et al. 2000,
Mazzotta et al. 2001, Ettori \& Fabian 2001, and references
therein). At $z=0$ the Coma--like halo is undergoing several merging
events with massive sub--groups, at different stages of advancement,
much like what is seen in the XMM observations of the real Coma cluster
(Briel et al. 2001).

\subsection{Definitions of observables}
The bolometric luminosity for a set of $N_{gas}$ gas particles is
defined as
\be
L_X\,=\,{m_{gas}\over \mu m_p}\,\sum_{i=1}^{N_{gas}}{\rho_i\over \mu
m_p}\,\Lambda_c(T_i)\,,
\label{eq:lx}
\ee 
where $m_p$ is the proton mass, $m_{gas}$ is the mass of a gas
particle (see Table \ref{t:simul}), $\rho_i$ and $T_i$ are the density
and temperature at the position of the $i$-th particle, respectively,
and $\mu=0.6$ is the mean molecular weight for a primordial gas
composition with 76\% mass provided by hydrogen. Assuming that the
main contribution to the $X$--ray emission is from free--free
bremsstrahlung, the cooling function turns out to be
$\Lambda_c(T)=1.2\times 10^{-24} (kT/{\rm{keV}})^{1/2}$ erg cm$^3$
s$^{-1}$ (e.g., Navarro et al. 1995).

Under the assumption of isothermal gas following the spherically
symmetric DM distribution, it can be shown that the bolometric
$X$--ray luminosity is
\be
L_X\,=\,\left({f_{gas}\over 3\mu m_p}\right)^2
\Delta_{vir}M_{vir}\bar\rho F(c)\,\Lambda_c(T)\,,
\label{eq:lth} 
\ee
(e.g., Eke et al. 1998) where $f_{gas}$ is the gas fraction
within the cluster, $\Delta_{vir}$ the overdensity corresponding to
virialization ($\simeq 178$ for $\Omega_m=1$; Eke, Cole \& Frenk
1996), $\bar\rho$ the average cosmic density and $M_{vir}$ the virial
cluster mass (i.e., the mass enclosed within the radius $R_{vir}$
within which the average density is $\Delta_{vir}\,\bar\rho$). Under
the above assumptions, mass and temperature are related according to 
\ba
k_BT & = & {1.38\over \beta}\,\left({M_{\rm vir}\over
10^{15}h^{-1}M_\odot}\right)^{2/3}\nonumber \\
& \times & \left[\Omega_m\Delta_{vir}(z)\right]^{1/3} \,(1+z)\,{\rm keV}\,,
\label{eq:mt}
\ea
(e.g., Eke et al. 1996). The $\beta$ parameter is defined as the ratio
of the specific kinetic energy of the collisionless DM
to the specific thermal energy of the gas,
\be
\beta\,=\,{\mu m_p\sigma_v^2\over k_BT}\,,
\label{eq:beta}
\ee
being $\sigma_v$ the one--dimensional velocity dispersion of DM
particles. For a Navarro-Frenk-White (1997, NFW) density profile,
$F(c)$ is a function of the concentration parameter $c$ only:
\be
F(c)\,=\,c^3\,{1-(1+c)^{-3}\over [\ln(1+c)-c/(1+c)]^2}\,.
\label{eq:fc}
\ee
For the sake of comparison between the prediction of eq.(\ref{eq:lth})
with results from simulations, we compute the concentration parameter
as a function of the halo mass for our choice of cosmological
parameters following the prescription of Eke, Navarro \& Steinmetz
(2001).

Besides the bremsstrahlung emission, line emission gives a
non--negligible contribution at temperatures $T<2$ keV. To account for
it and correctly compare our results to observations, we use the
Raymond--Smith (1977) code in the XSPEC package to estimate the
emissivity of a plasma with metallicity $Z=0.3Z_\odot$.

We define the mass--weighted temperature of the cluster as
\be
T_{mw}\,=\,{1\over N_{gas}}\,\sum_{i} T_i\,,
\label{eq:tmw}
\ee 
where the sum is over all the particles falling within the cluster
virial radius. A more useful quantity for comparison with observations
is the emission--weighted temperature, which is defined as 
\be
T_{ew}\,=\,{\sum_{i} \rho_i\,T_i^{3/2}\over \sum_{i}
\rho_i\,T_i^{1/2}}\,,
\label{eq:tew}
\ee
This definition strictly holds only for bremsstrahlung emissivity. We
verified that final values of $T_{ew}$ are essentially unchanged
if we account for the contribution from metal lines.

Finally, we define the gas entropy carried by the $i$-th particle as
\be
s_i\,=\,{kT_i\over n_{e,i}^{2/3}} {\rm{keV~cm^2}}\,,
\label{eq:entr}
\ee
where $n_{e,i}$ is the electron number density associated with that
particle. 

In Table \ref{t:glob} we give the values of bolometric luminosity,
mass and emission--weighted temperatures for the simulated
structures and for the different pre--heating schemes (see below).

\begin{table*}
\centering
\caption[] {Quantifying the simulated ICM at $z=0$. Column 1 shows
the type of
the run and corresponding label. For the runs with SN
feedback $\eta$ represents the fraction of the total energy released
by SN which is thermalized into the ICM. For the S-50 and S-100 runs,
$S_{\rm fl}$ indicates the value of the entropy floor set at
$z=3$. $L_{br}$ is the $X$--ray luminosity from bremsstrahlung
emission 
($10^{43}\lum$); $L_{RS}$ is the $X$--ray luminosity using a
Raymond--Smith code to account for the contribution from line
emission, assuming $Z=0.3Z_\odot$ for the average gas metallicity
$10^{43}\lum$); $T_{mw}$ and $T_{ew}$ are mass--weighted and
emission--weighted temperature within $R_{vir}$ (keV). The asterisks
mark the simulations for which only the LR runs are available.}
\tabcolsep 3pt
\begin{tabular} {lcccccccccccccccc}
         & \multicolumn{4}{c}{Coma} & \multicolumn{4}{c}{Virgo} & \multicolumn{4}{c}{Fornax} & \multicolumn{4}{c}{Hickson} \\
Run type & $L_{br}$ & $L_{RS}$ & $T_{mw}$ & $T_{ew}$ & $L_{br}$ & $L_{RS}$ & $T_{mw}$ & $T_{ew}$ & $L_{br}$ & $L_{RS}$ & $T_{mw}$ & $T_{ew}$ & $L_{br}$ & $L_{RS}$ & $T_{mw}$ & $T_{ew}$ \\ ~ \\
\noalign{\smallskip}
Grav. heating (GH)       & 139 & 160 & 5.1 & 5.8 & 35.2 & 42.0 & 1.99 & 2.70 & 6.52 & 10.7 & 0.69 & 0.95 & 1.61 & 3.87 & 0.39 & 0.60 \\
SN feedback ($\eta=0.1$, SN-0.1) &  &  &  &  &  &  &  &  &  &  & &
& 0.73$^*$ & 1.71$^*$ & 0.38$^*$ & 0.60$^*$ \\
SN feedback ($\eta=1$, SN-1)   & & & & & 32.0 & 37.7 & 2.01 & 2.75 & 3.63 & 5.65 & 0.68 & 1.02 & 0.45 & 1.02 & 0.39 & 0.63 \\ 
SN feedback ($\eta=4$, SN-4) &  &  &  &  &  &  &  &  &  &  & &
& 1.8e-2 & 3.8e-2 & 0.46 & 0.62 \\
$S_{\rm fl}=50$ keV cm$^2$ (S-50) & 103 & 117 & 5.2 & 7.2 & 14.3 & 16.7 & 2.07 & 3.00 & 0.38 & 0.67 & 0.60 & 0.94 & 0.04 & 0.10 & 0.39 & 0.58 \\
$S_{\rm fl}=100$ keV cm$^2$ (S-100)   &  &  &  &  & 3.2 & 3.8 & 1.99 & 2.77 & 9.9e-2 & 0.18 & 0.58 & 0.90 & 1.1e-2 & 2.4e-2 & 0.40 & 0.58 \\
Cooling SN-1 &  &  &  &  &  &  &  &  &  &  & &
& 3.1e-2 & 6.1e-2 & 0.41 & 0.81

\end{tabular}
\label{t:glob}
\end{table*}

\subsection{Effects of resolution}
It is generally regarded that a few thousands of DM particles within
the virial radius are required to describe global quantities, such as
total virial mass $M_{vir}$ and velocity dispersion $\sigma_v$, and as
many gas particles are also sufficient to correctly estimate the ICM
temperature. However, a resolution higher by about one order of
magnitude is required to correctly describe the $X$--ray luminosity:
since this quantity depends on the square of the local gas density,
its correct computation requires the details of cluster structure in
the innermost regions and gas clumping to be accurately resolved
(e.g., Anninos \& Norman 1996, Navarro et al. 1995, Eke et al. 1998,
Bryan \& Norman 1998, Lewis et al. 2000).

Resolution is even more an issue if one wants to  properly resolve
the internal structure of the ICM, including shock patterns of the
infalling gas, which carry information on the physics of the diffuse
medium.  The higher the resolution, the larger the number of
substructures surviving within the virial radius of the main cluster
halo. The largest sub--halos are able to keep a fraction of their gas
and can carry it down to the center of the main halo, possibly
shocking the gas and raising the central entropy. The $X$--ray
luminosity profiles also depend on the underlying matter profile,
for
which about a hundred thousand particles are necessary to avoid
underestimating its central density (Moore et al. 1998).

\begin{figure} 
\centerline{ \hbox{
\psfig{file=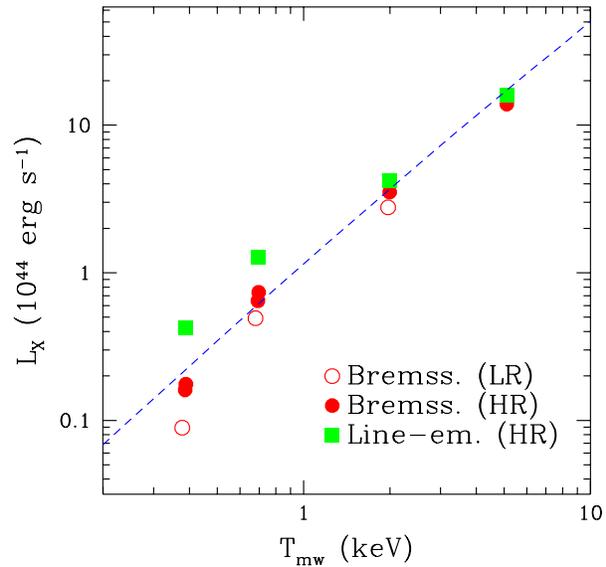,height=8.5cm,width=8.5cm,angle=0} } }
\caption{The effect of numerical resolution and line--emission on the
relation between $X$--ray luminosity and mass--weighted temperature
for the GH runs. Open and solid circles refer to pure bremsstrahlung
luminosity for the set of low--resolution and high--resolution runs,
respectively (see Table 1). Squares are for luminosity computed by
accounting for the contribution from line emission, estimated from a
Raymond--Smith code assuming $Z=0.3Z_\odot$ for the global ICM
metallicity. For the Fornax and Hickson runs, the almost overlapping
lower and upper circles are for softening parameter $\epsilon =5$ and
2.5 kpc, respectively (see text). The dashed line is the prediction of
the model described in the text, based on a bremsstrahlung cooling
function, an NFW profile for the dark matter halo, with concentration
parameter appropriate for the simulated cosmology, and an $M_{\rm
vir}$--$T$ relation given by eq.(\ref{eq:mt}) with $\beta=1$.}
\label{fi:lt_res} 
\end{figure}

In Figure \ref{fi:lt_res} we compare the relation between $X$--ray
luminosity and mass--weighted temperature, $T_{mw}$, for the HR and LR
runs with gravitational heating (GH) only.  Increasing the resolution
has a non--negligible effect on the estimated $X$--ray luminosity,
while, as expected, has only a marginal effect on the mass--weighted
temperature. The $L_X$ value for the HR run of the Virgo cluster is
$\sim 25\%$ higher than for the LR run, the difference increasing to
$\sim 50\%$ for the Fornax group and to $\sim 100\%$ for the Hickson
group. We also verified that decreasing the softening by a factor two
for the Fornax and Hickson runs, while keeping the mass resolution
fixed, increases $L_X$ by about 10\%, thus showing that the adopted
spatial resolution is adequate to resolve all the structures which are
responsible for the $X$--ray emission. In fact, at the highest
resolution of our runs, the number of massive subhalos (i.e., with
circular velocity larger than one--tenth of that of the main halo) is
likely to have converged (Ghigna et al. 2000). We expect spatial
resolution to be even less of an issue when considering simulations
with extra heating (see below). Indeed, in this case the gas is put on
a higher adiabat and, therefore, does not follow the small--scale
potential wells which are characterized by a virial temperature of a
few tens of keV.

\begin{figure*} 
\centerline{ \hbox{
\psfig{file=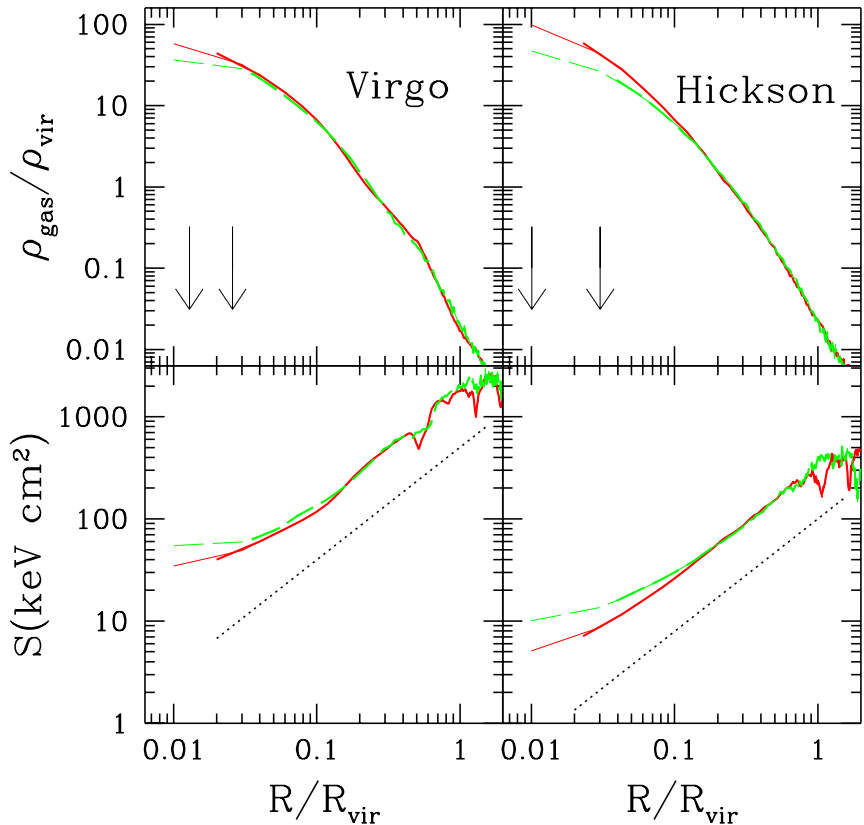,height=12cm,width=12cm,angle=0} } }
\caption{Effect of resolution on the profiles of gas density (upper
panels) and specific entropy (lower panels) for the GH version of the
Virgo and Hickson simulations.  In each panel continuous and dashed
curves correspond to the HR and LR runs, respectively. Curves become
thinner at the radius containing 100 gas particles. The vertical
arrows mark the scale corresponding to three times the value of the
Plummer--equivalent force softening for the LR and HR runs (see Table
1), therefore close to the actual spline softening length. The dotted
lines in the lower panels show the $S\propto R^{1.1}$ entropy profile
predicted by the semi--analytical model by TN01.}
\label{fi:profs_res} 
\end{figure*}

We stress that the HR runs predict a $L_X$--$T$ relation which agrees
well, both in slope and normalization, with the prediction of the
scaling model of eq.(\ref{eq:lth}) for a pure bremsstrahlung cooling
function. We consider this agreement as a convincing indication that
the HR runs provide a correct description of the gas distribution,
which requires using at least $\simeq 5\times 10^4$ particles within
the virial radius. This result is in agreement with that from
resolution studies involving the collisionless component only (e.g.
Moore et al. 1998), which established the minimum number of particles
required to correctly model the central profile of dark matter halos.
Fig. \ref{fi:lt_res} also highlights a common problem of simulations
of large cosmological volumes at {\it fixed mass resolution}: as
resolution becomes worse for smaller masses, $L_X$ becomes
underestimated. For instance, taking the same mass and spatial
resolution for the Virgo, Fornax and Hickson runs would produce
spurious steepening of the \lt relation. Finally, while a pure
bremsstrahlung emissivity is a good approximation at $T\magcir 2$ keV,
the effect of line emission is shown to become important for
smaller/cooler systems. The net effect of properly accounting for them
is that of flattening the \lt relation, thus further increasing the
discrepancy with respect to observations at the scale of poor clusters
and groups.

In Figure \ref{fi:profs_res} we show the effect of the resolution on
the profiles of gas density and mean entropy per particle. There is an
overall good convergence of these quantities, at the radii
encompassing at least 100 gas particles (tick curves), with slight
deviations at somewhat larger scales for the Hickson run. This number
of particles corresponds to about three times the number of neighbors
over which the SPH smoothing length is estimated. Therefore, the
radius containing this many particles indicates the smallest scale
which is resolved by the SPH kernel with the given softening. On
smaller scales the gas density tends to be somewhat underestimated and
so is the $X$--ray luminosity, while two--body heating causes some
flattening of the entropy profile (e.g., Steinmetz \& White 1997,
Yoshida et al. 2001). As mass and spatial resolutions are increased,
gas and entropy profiles steepen in the innermost regions. The absence
of an entropy core agrees with predictions from analytical models
based on gravitational gas heating and spherical accretion: gas
residing in the central cluster regions has been accreted at the very
beginning and, therefore, never significantly shocked, thus preserving
its initial low entropy.

\begin{figure} 
\vspace{-1.7truecm}
\centerline{\hbox{
\psfig{file=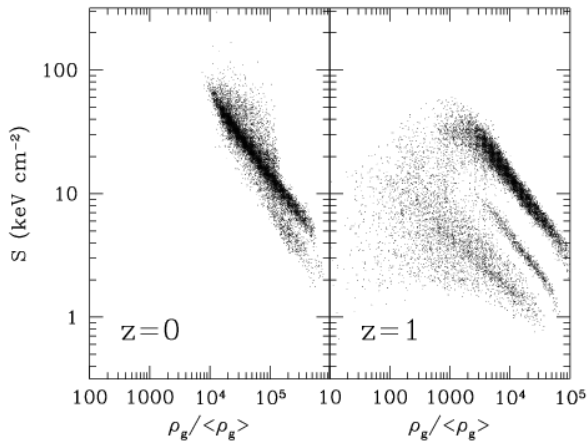,width=9.5cm} } }
\vspace{-1.8truecm}
\caption{Gas entropy vs. overdensity for particles of the Fornax group
which fall within $0.1R_{\rm vir}$ at $z=0$ (left panel) and traced
back to $z=1$ (right panel).}
\label{fi:entr01} 
\end{figure}

Besides extra heating, a further possibility to increase the entropy
level in central cluster regions could be by anisotropic accretion of
gas which has been previously gravitationally shocked by large--scale
filaments. In order to verify this possibility, we identify at $z=0$
gas particles in the central region of a halo and trace them back at
higher redshift. As an example, we show in Figure \ref{fi:entr01} the
density--entropy scatter plot for the gas particles falling at $z=0$
within $0.1R_{vir}$ in the HR Virgo simulation, and their
corresponding distribution at $z=1$. In the left panel these particles
define a darker strip, which correspond to the cluster core, while the
lighter strip at lower entropy witnesses the presence of a merged
subclump which still preserves its identity (see also Fig.
\ref{fi:gasmaps}). At $z=1$ these same particles define several
individual subclumps, which by the present time have merged together
at the center of the main halo. Such groups at $z=1$ correspond to
lower entropy structures, in line with the expectation that the gas
within smaller structures undergoes weaker shocks. Gas possibly
accreted along filaments would correspond at $z=1$ to structures with
density contrast $\delta\sim 10$ containing shocked gas with $S\magcir
50$ keV cm$^2$. Actually, there is no trace of such gas particles in
the right panel of Fig. \ref{fi:entr01}. This shows that gas accreted
along filaments does not penetrate efficiently into the cluster
central region and contribute to increasing the entropy
there. 

\section{Beyond gravitational heating}
\subsection{Entropy floor}
Our  first scheme for non--gravitational heating is based on setting a
minimum entropy value at some pre--collapse redshift (e.g. Navarro et
al. 1995, Bialek et al. 2001, TN01). For gas with
local electron number density $n_e$ and temperature $T$, expressed in
keV, at redshift $z$, we define the entropy as
\be 
S\,=\,{T\over n_e^{2/3}}\,=\,\left[{f_{\rm bar}\over
m_p}\,{1+X\over 2}\,\bar\rho(z_h)\,(1+\delta_{\rm
g})\right]^{-2/3}\,T~{\rm keV\,cm^2}\,,
\label{eq:entr1}
\ee
where $\bar\rho(z)=\bar\rho_0(1+z)^3$ is the average cosmic matter
density at redshift $z$, $\delta_{\rm g}$ the gas overdensity and $X$
the hydrogen mass fraction. We choose two values for this entropy floor,
$S_{\rm fl}=50$ and 100 keV cm$^2$, that bracket the observed values
in small groups and clusters (Ponman et al. 1999, Lloyd--Davis et
al. 2000).

We assume $z_h=3$ for the reference heating redshift, since it is
close to the epoch at which sources of heating, like SN or AGNs, are
expected to reach their maximum activity (see below).  At $z=3$, we
select all the gas particles with overdensity $\delta_{\rm g}>5$, so
that they correspond to structures which have already undergone
turnaround. After assuming a minimum floor entropy, $S_{\rm fl}$, each
gas particle having $s_i<S_{\rm fl}$ is assigned an extra thermal
energy, so as to bring its entropy to the floor value. We estimate the
amount of energy injected in the ICM in these pre--heating schemes by
selecting at $z=0$ all the gas particles within the virial radius and
tracing them back to $z=3$. We find that taking $S_{\rm fl}=50$ keV
cm$^2$ amounts to give an average extra heating energy of $E_h={3\over
2}T_h\simeq 1.4$ keV/part for particles that end up within the virial
radius of the Fornax and Hickson groups at $z=0$, and $E_h\simeq 0.9$
keV/part for the Virgo cluster and $E_h\simeq 0.8$ keV/part for the
Coma cluster. Such values are twice as large for $S_{\rm fl}=100$ keV
cm$^2$.

In the semi--analytical model by TN01, the gas is assumed not to have
undergone any significant gravitational shock heating before $z_h$, so
that eq.(\ref{eq:entr1}) actually gives the extra energy to be
provided to the gas particles to bring them to the appropriate
adiabat. The hierarchical clustering scenario, instead, predicts that
a significant amount of non--linear structures already exist at
$z_h=3$, with gas particles raised by gravitational shocks to an
entropy level already higher than the floor to be set.  In our
heating scheme, the thermal energy of such particles is left
unchanged.

This heating scheme can be defined as an external one, in the sense
that it is also targeted at particles not belonging to collapsed
structures. In principle this may violate constraints on the entropy
of the diffuse high--$z$ IGM, as inferred from observations of
Ly--$\alpha$ absorption systems. If we identify a Ly--$\alpha$ system
as a structure with typical gas overdensity $\delta_{\rm g}\simeq 30$,
it has at $z=3$ a temperature $T\simeq 2.5 \times 10^{-3}$ keV, as
determined by the condition of ionization equilibrium with the
UV--background (e.g. Haardt \& Madau 2001). This corresponds to a
typical entropy $S\simeq 5$--10 keV cm$^{-2}$, thus about an order of
magnitude smaller than that relevant for ICM heating. Therefore, an
external heating such as the one imposed in our simulations would
destroy Ly--$\alpha$ absorbers. In this sense, setting an entropy
floor must just be considered as a approach guided by ICM
phenomenology rather than motivated by general astrophysical
arguments.

Even in this simple heating scenario, a further parameter is
represented by the epoch at which the entropy floor is created.
As the heating redshift is changed, two opposite effects compete in
determining the energy budget required to create a given entropy
floor. On the one hand, the higher $z_h$, the smoother the particle
distribution, the smaller the number of particles at high $\delta_{\rm
g}$, which require a lot of heating to increase their entropy to the
desired level. On the other hand, the higher $z_h$, the higher the
overall cosmic mean density $\bar \rho(z)$, the smaller the amount of
heating energy needed at fixed $\delta_g$.  We check the effect of
changing $z_h$ by also running low--resolution simulations of the
Fornax group with $z_h=1$, 2 and 5. We will discuss in the
following the effect of changing $z_h$ on the final results. The
heating energies corresponding to the different $z_h$ are shown in the
top panel of Figure \ref{fi:zh}.

\subsection{Energy feedback from supernovae}
This pre--heating scheme is based on computing the star--formation
rate within clusters as predicted by a semi--analytic model of galaxy
formation, a technique first introduced by White \& Frenk (1991),
Kauffmann, White \& Guiderdoni (1993) and Cole et al. (1994), and
subsequently adopted and developed by several authors (e.g. Somerville
\& Primack 1999, Wu, Fabian \& Nulsen 2000, Cole et al. 2000). Here we
use a variation of the scheme described by Menci \& Cavaliere
(2000), and we refer to this paper for a detailed description of the
method. 

In our approach, the merging history of dark--matter halos, having the
same mass of the simulated ones, is predicted according to the
Extended Press--Schechter theory (EPST, e.g. Lacey \& Cole 1993). The
processes of gas cooling, star formation and stellar feedback within
galaxy--sized structures are described by means of a suitable
parametrization of the corresponding star formation rate and 
reheated mass, which is ejected back into the hot diffuse medium.
The free parameters of the model are chosen so as to
reproduce observed properties of the local galaxy population
(Tully--Fisher relation, B-- and K--band luminosity functions,
disk--sizes). 

\begin{figure} 
\centerline{ \hbox{
\psfig{file=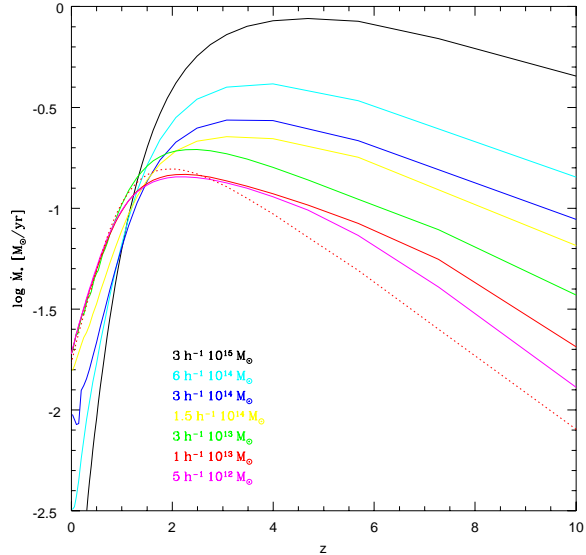,height=8cm,width=8cm,angle=0} } }
\caption{The integrated star formation rate for the condensations
ending up in a present-day structure of mass $M_0$ is plotted as a
function of redshift for the different values of $M_0$ shown in the
labels. For comparison, the dotted curve shows the cosmic star
formation rate predicted by our semi--analytical scheme.}
\label{fi:sfr} 
\end{figure}

In Figure \ref{fi:sfr} we show the integrated star formation history
$\dot m_*(z,M_0)$ of all the condensations which are incorporated into
a structure of total mass $M_0$ by the present time. The curves in
Fig. \ref{fi:sfr} correspond to $M_0$ values, which span the whole
range from galaxies to rich clusters. As $M_0$ grows the local star
formation rate (SFR) decreases, while the value at $z\magcir 2$
increases and its peak is attained at higher redshift. This is
expected in the hierarchical clustering picture, since more massive
structures are originated from more biased regions, where initially
the star formation rate was higher. The large consumption of cold
baryons in the progenitors of $M_0$ at large $z$ results in a
smaller amount of available star-forming gas at small $z$, yielding
the sharply declining $\dot m_*$ at $z\rightarrow 0$.

After assuming a Salpeter initial mass function (Salpeter 1955), the
SFRs are then used to derive the SN heating appropriate to each system
that we simulate, based on its virial mass. During the evolution of
each halo, this energy is shared among all the gas particles having
$\delta_g> 50$. This overdensity threshold, which roughly corresponds
to the density contrast at the virial radius, guarantees that gas
heating takes place inside virialized regions.  We verified that the
final results do not change if only gas particles with $\delta_g>500$
are heated. Under the extreme assumption that all the energy released
by SN is thermalized into the ICM (i.e. $\eta=1$ for the SN
efficiency) this scheme dumps a total amount of about 0.35 keV per gas
particle. For the Hickson group we also run two more simulations with
$\eta=0.1$ and $\eta=4$. In the first case, the energy budget is so
small that the final results are essentially indistinguishable from
the GH run (see Table \ref{t:glob}) and, therefore, we will not
comment further on this low--efficiency case.  As for the second case,
it corresponds to an amount of extra energy similar to that of the
S--50 entropy floor. This allows us to check whether the final results
depend on the way a fixed amount of energy is dumped into the diffuse
medium. Such a large amount of energy can be interpreted as associated
with any heating source whose evolution follows that of the SFR. This
is likely the case for AGNs, whose emissivity per unit volume has been
suggested to evolve according to the SFR (e.g. Cavaliere \& Padovani
1989, Boyle \& Terlevich 1998, Franceschini et al. 1999).

A word of caution should be mentioned here about our implementation of
SN heating. We assume that all SNe release a fixed amount of energy,
$E_{SN}=10^{51}$ ergs. In the so--called hypernovae scenario, energies
larger by at least one order of magnitude may be expected
(e.g. Loewenstein 2001).

Furthermore, in our scenario only Type II SN are included, while Type
Ia SN may in principle give a significant contribution to the ICM
energetics (e.g., Pipino et al. 2002). Also, most of our simulations
assume no radiative losses in SN explosions, so that the released
energy is all thermalized in the ICM, which is a reasonable
approximation if most of the SN explode in a hot and rarefied
medium. Finally, the feedback used in the semi--analytical modeling of
galaxy formation is not the same as that adopted for heating gas
particles in the simulations. For all these reasons, we tend to
consider our SN heating scheme as a realistic but approximate recipe
to estimate the amount of energy expected from a stellar population,
rather than a self--consistent approach to including the effect of SN
explosions on the ICM energetics.

\section{Results}

\begin{figure*} 
\centerline{ \hbox{
\psfig{file=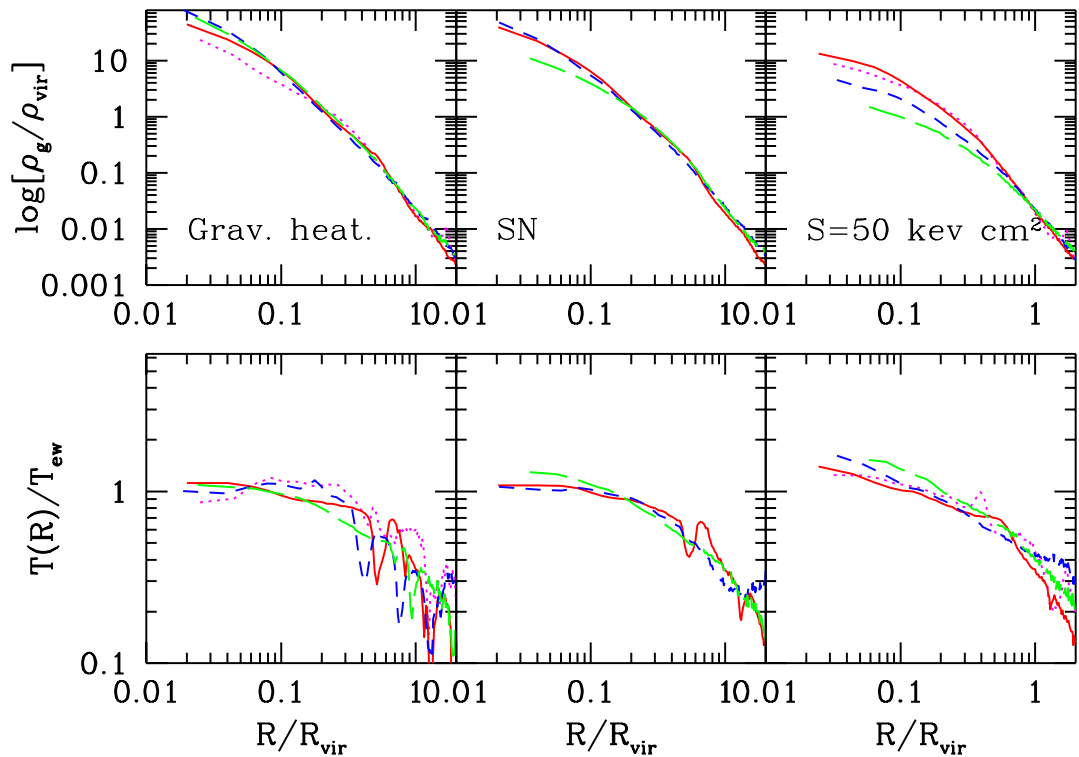,height=15cm,width=15cm,angle=0} } }
\vspace{-4.5truecm}
\caption{Profiles of gas density in units of the virial density
(upper
panels), and temperature in units of $T_{ew}$ (lower panels) at
$z=0$. Left, central and right panels are for the GH, SN and S-50
runs, respectively. Dotted, solid, short--dashed and long--dashed
curves refer to the Coma, Virgo, Fornax and Hickson runs,
respectively.}
\label{fi:profs} 
\end{figure*}

\subsection{Gas--density and temperature profiles}
A unique signature of the lack of self--similarity in galaxy clusters
is provided by the different profiles of gas density observed for
clusters of different temperatures (e.g. Ponman et al. 1999,
Lloyd--Davis et al. 2000), with colder systems having shallower
internal profiles. In the upper panels of Figure \ref{fi:profs} we
show how extra heating can affect the gas profiles. In the absence of
any extra heating, all the simulated structures display the same gas
density profile. This result agrees with the expectation of the
self--similar scaling paradigm and confirms that non--linear
gravitational processes, like accretion shocks, do not introduce any
characteristic scale. The only structure slightly deviating from
self--similarity in the GH runs is the Coma cluster. Looking at its
evolution, it turns out that this structure underwent a very recent
major merger event, which redistributed the gas in the central
regions. Therefore, its slightly softer gas profile is just the
signature of this recent merger.

Self--similarity is broken as additional heating is included. SN
feedback has some effect only on the gas profile for the Hickson
group, our less massive halo.  This is in line with the expectation
that ICM thermodynamics ought to be changed only when the
non--gravitational heating is comparable to the gravitational one. In
fact, SN heating provides $\sim 1/3$ keV per gas particle, which
is similar to the mass--weighted temperature of the Hickson group.
As for the S-50 runs, the entropy floor corresponds to a large enough
heating that only the Coma cluster remains almost unaffected. This
result is in line with that based on the shape
of entropy profiles presented in Paper I.

\begin{figure*} 
\centerline{ \hbox{
\psfig{file=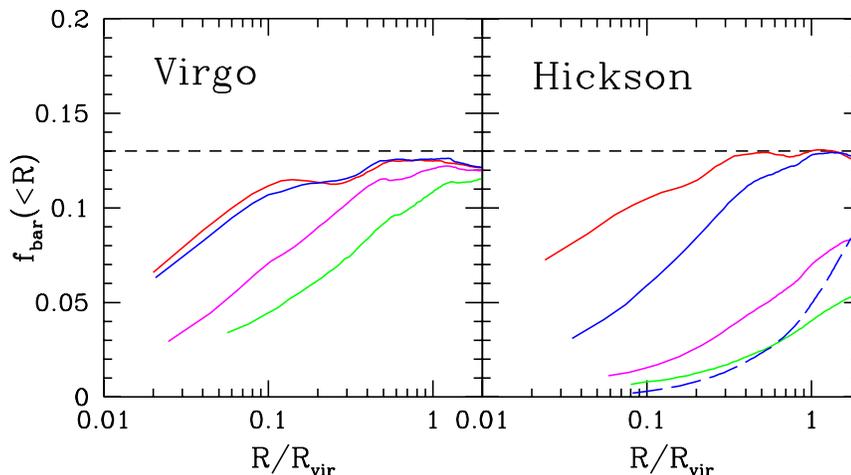,height=12cm,width=12cm,angle=0} } }
\vspace{-5.5truecm}
\caption{Profiles of the baryon fraction within a given radius for
Virgo and Hickson runs. Gravitational heating, SN feedback, and
entropy floors with $S=50$ and 100 keV cm$^2$ are shown from upper to
lower solid curves. The dashed curve for the Hickson run indicates the
SN heating scheme with efficiency raised to 400\% (SN-4). The horizontal
dashed line indicates the cosmic baryon fraction.}
\label{fi:fbar} 
\end{figure*}

A relevant quantity which is directly related to the gas profile is
the baryon fraction, $f_{\rm bar}$, within a given radius.  The
measurement of the baryon fraction in clusters, in combination with
constraints on $\Omega_b$ from primordial nucleosynthesis, is
considered as one of the fundamental tests for the density parameter
$\Omega_m$ (e.g., White et al. 1993, Evrard 1997, Ettori 2002). We
show in Figure \ref{fi:fbar} the radial dependence of $f_{\rm bar}$
for the Virgo and Hickson runs with the different heating schemes.  If
gas is only subject to gravitational heating, the cosmic baryons
fraction is recovered at about 0.3--0.4 $R_{\rm vir}$, which is
comparable to the effective scale where $f_{\rm gas}$ is determined
from $X$--ray data. As for the Virgo cluster, the S-50 scenario would
lead only to a $\sim 10$--15\% underestimate of the correct value at
about half virial radius. The situation is quite different for the
Hickson group, whose gas fraction is diminished out to the virial
radius by heating from the entropy floors. This provides a warning
about the determination of $f_{\rm gas}$ and the inference of
$\Omega_m$ based on poor clusters and groups. The heating from the SN
scheme with 400\% efficiency (SN-4) produces a $f_{\rm gas}$ profile
quite similar to the S-100 scheme within the virial radius. Quite
interestingly, the S-100 scheme involves an extra--heating energy,
$E_h\simeq 2.8$ keV, which is almost twice as large than for the
former scheme. This indicates that a redshift-modulated heating is
more efficient, in terms of the required energy budget, than an
impulsive pre--collapse heating.

As for the three-dimensional temperature profiles (lower panels of
Fig. \ref{fi:profs}), the gas looks almost isothermal, in the
absence of
any extra heating, out to $\sim 0.2R_{\rm vir}$, while a fairly steep
gradient appears at larger radii. Dips in the $T$--profiles
indicate the presence of merging structures, whose lower virial
temperature brings colder gas in the main body of the simulated
system, before being stripped by ram pressure and thermalized. Such
structures are progressively washed out in the pre--heated runs, as a
consequence of the smaller amount of gas that subgroups are able to
keep in such cases. The extra heating also steepens the
temperature profiles in the internal cluster regions, with
a more pronounced gradient for smaller systems. An increase of gas
temperature in the central regions is the consequence of suppressing
gas
density while keeping unchanged the external gas pressure.

\begin{figure} 
\centerline{ \psfig{file=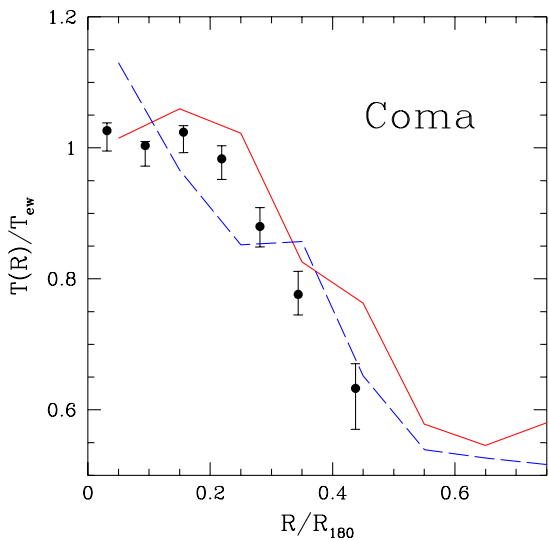,width=8cm} }
\caption{The projected temperature profile for the GH and S-50 Coma
runs (solid and dashed lines, respectively), compared with the average
temperature profile for the set of 21 clusters observed with
Beppo--SAX and analyzed by De Grandi \& Molendi (2002). Simulation
results are obtained by averaging the profiles obtained by projecting
along three orthogonal directions. The cluster--centric radius is
expressed in units of $R_{180}$, which is defined as the radius
encompassing the density $180\rho_{\rm crit}$.}
\label{fi:tproj}
\end{figure}

In the last few years, observational data have reached good enough
precision to allow for spatially--resolved spectroscopic observations
of the ICM for fair number of galaxy clusters. Yet, a general
agreement about the temperature profiles in the external cluster
regions, not affected by cooling, has still to be reached (e.g.,
Markevitch et al. 1998, White 2000, Irwin \& Bregman 2001).  Based on
Beppo--SAX observations, De Grandi \& Molendi (2002) analyzed
projected temperature profiles for a set of 21 clusters. They found
the gas to be isothermal out to $\sim 0.2 R_{180}$, with a fairly
steep negative gradient at larger radii.  In Figure \ref{fi:tproj} we
compare result from our simulations to the average profile by De
Grandi \& Molendi (2002). Since their clusters are all quite hot
($T\magcir 4$ keV), we use only results from the Coma run. The
simulation results are obtained by averaging over the temperature
profiles projected along three orthogonal directions.  The simulation
produces a fairly steep gradient at $R\magcir 0.2 R_{180}$, quite
similar to the observed one. Forthcoming data from Chandra and
Newton--XMM (e.g. Arnaud et al. 2001, Allen, Schmidt \& Fabian 2001)
will certainly refine the determination of temperature profiles in
central cluster regions, where cooling is also expected to play a
significant role. There is no doubt that such observational advances
will challenge the ability of numerical simulations to treat the
relevant physical processes.

\subsection{The mass--temperature relation}
The $M$--$T$ relation represents a key ingredient when comparing
observational data on the cluster $X$--ray temperature function with
the mass function predicted by cosmological models. Quite recently,
the different results obtained by different authors on the
power--spectrum normalization, $\sigma_8$, from cluster XTF and XLF
have lead to a discussion on the correct $M$--$T$ relation to be used
(e.g. Pierpaoli, Scott \& White 2001, Borgani et al. 2001b, Seljak
2002, Viana, Nichol \& Liddle 2002, Ikebe et al. 2002). In past years,
different analyses have shown that eq.~(\ref{eq:mt}) provides a good
fit to results of simulations including only gravitational heating,
with values of $\beta$ ranging within the interval $0.9\mincir
\beta\mincir 1.3$ (e.g. Evrard, Metzler \& Navarro 1996, E96
hereafter, Bryan \& Norman 1998, Frenk et al. 1998). In particular,
E96 pointed out that temperature measurements provide a quite accurate
determination of the mass at radii where the mean cluster density is
500--$2500\,\rho_{\rm crit}$.

\begin{figure*}
\centerline{ \hbox{
\psfig{file=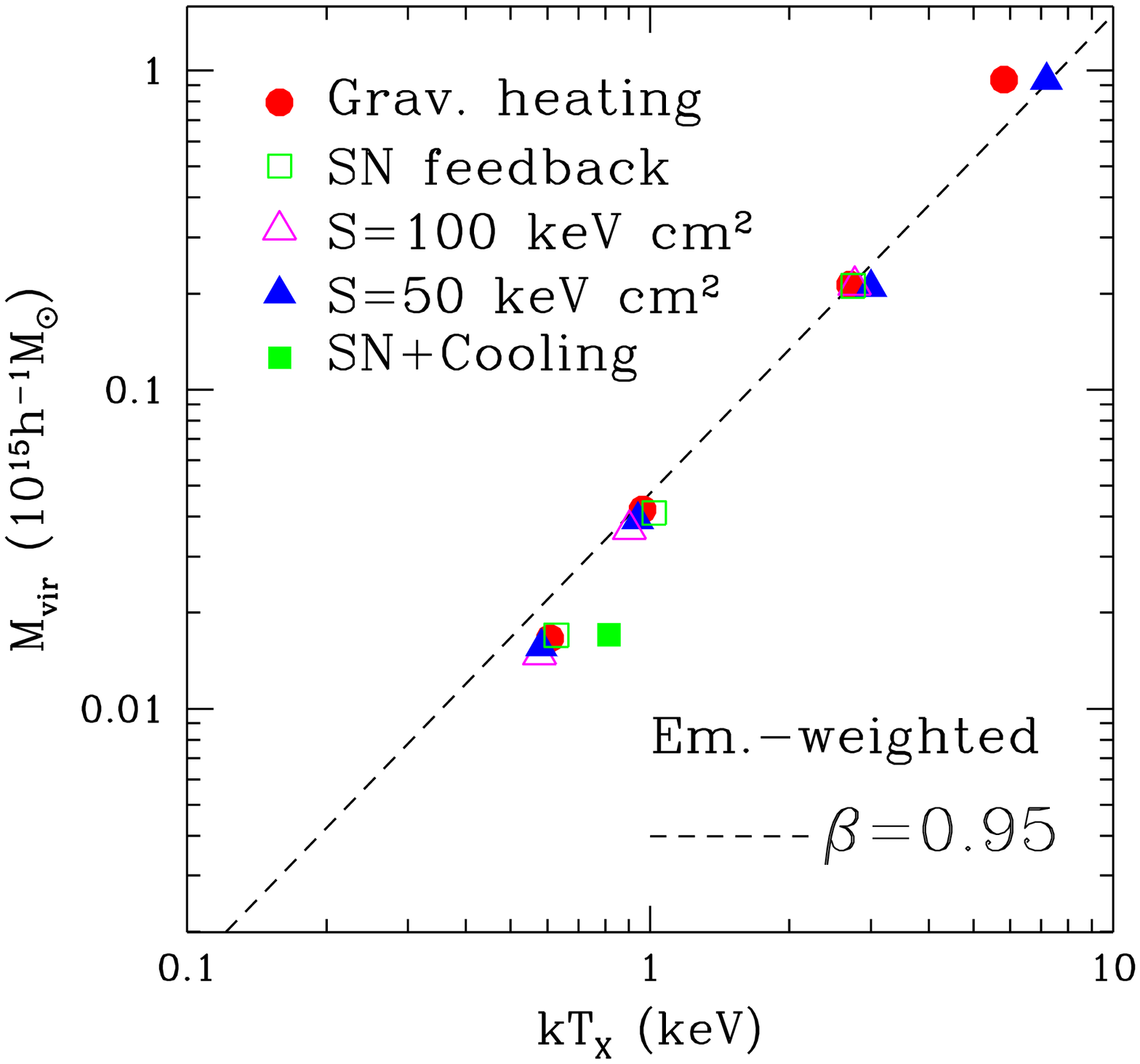,width=8cm} \psfig{file=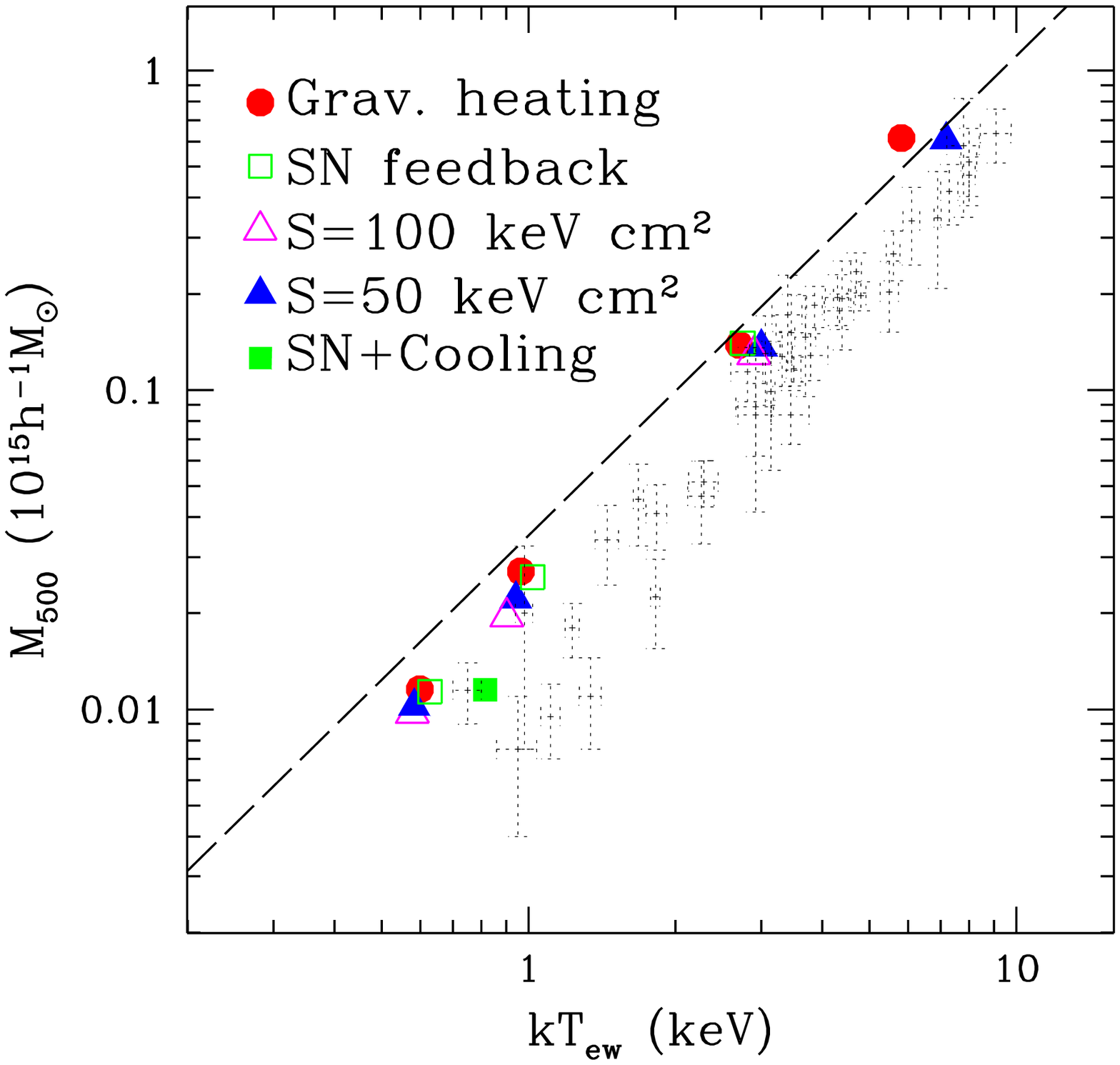,width=8cm}}}
\caption{Left panel: the relation between total virial mass and
emission--weighted temperature for simulated clusters. The dashed line
shows the relation of eq.(\ref{eq:mt}), with normalization as given by
the reported $\beta$ values. Right panel: the relation between mass
estimated at $r_{500}$ and emission--weighted temperature. Dotted
crosses are the observational results by Finoguenov et al. (2001). The
dashed line shows the best--fitting relation found by Evrard et
al. (1996).}
\label{fi:mtvir} 
\end{figure*}

Besides investigating the effect of non--gravitational heating on the
$M$--$T$ relation, the excellent resolution of our simulations allows
to better resolve the structure of cluster cores, which is relevant
for determinations of the emission--weighted temperature.  Results on
the $M_{\rm vir}$--$T_{mw}$ relation from our simulations (left panel
of Figure \ref{fi:mtvir}) closely follow the relation of hydrostatic
equilibrium of eq.(\ref{eq:mt}) with a very small scatter and almost
no dependence on non--gravitational heating. Although extra heating
adds internal energy to the gas particles, they quickly get back in
hydrostatic equilibrium, with their temperature determined by the
depth of the gravitational potential well which is established by the
dynamically dominant DM component.

As for observational data,  the $M$--$T$ relation
has been measured  in the last few years by different groups using
ASCA (Horner et al. 1999, Nevalainen et al. 2000, Finoguenov et
al. 2001, F01 hereafter) and Beppo--SAX data (Ettori, De Grandi \&
Molendi 2002). Such analyses consistently find that: {\it (a)}
$M\propto T^{3/2}$ for $T\magcir 4$ keV, but with a normalization
significantly lower than that found by E96 from simulations
(cf. Ettori et al. 2002); {\it (b)} a steeper slope for colder
systems, possibly interpreted as an effect of pre--heating.  We
compare in the right panel of Figure \ref{fi:mtvir} data on the
$M_{500}$--$T_{ew}$ relation by F01, which include data on systems
down to $T_{ew}\mincir 1$ keV, to results from our simulations and to
the relation found by E96. The somewhat lower normalization with
respect to the E96 results on the group scale is likely to be due to
our improved resolution.  However the effect of extra heating is at
most marginal and not sufficient to reconcile simulations to data,
thus indicating that some other physical process should be at work in
establishing the $M$--$T$ scaling.

Finoguenov et al. (2001, F01 hereafter) suggest that the difference
between observed and simulated $M$--$T$ relation is due to the
combined effect of pre--heating and the effect of formation redshift
on the temperature of the system. However, our simulations show that
pre--heating has a minor impact on this relation. As for the effect of
formation redshift, it may introduce a bias in the definition of the
observational data set: observations could tend to select fairly
relaxed systems, which formed at higher redshift and, therefore, are
characterized by a somewhat higher temperature at a fixed mass
(e.g. Kitayama \& Suto 1996, Voit \& Donahue 1998). Our simulations
define an $M$--$T$ relation with a small scatter, thus suggesting that
differences in the formation epoch or differences in the current
dynamical status among systems should have a small effect. A larger
set of simulated clusters would be required to properly address this
point. 
Ettori et al. (2002) detect a segregation in the $M$--$T$ relation for
cooling--flow and non cooling--flow clusters, the latter being
characterized by a larger scatter.

TN01 showed that their ICM model, which incorporates the effects of
pre--heating and cooling, reproduces the observed $M$--$T$
relation. They also find that the predicted relation is weakly
sensitive to the value of the pre--collapse entropy floor. This
suggests that cooling should be responsible for the lower
normalization of the relation, through the steepening of the
temperature profiles in cluster central regions. The effect of cooling
on the $M$--$T$ relation will be further discussed in Section 5 below.

\subsection{The luminosity--temperature relation}
The observed relation between bolometric luminosity and temperature is
considered a standard argument against the self--similar behavior of
the ICM. Bremsstrahlung emissivity predicts $L_X\propto M\rho_{\rm
gas}T^{1/2}$. Therefore, as long as clusters of different mass are
scaled versions of each other, then the $M$--$T$ scaling from
hydrostatic equilibrium gives $L_X\propto T_X^2(1+z)^{3/2}$ or,
equivalently $L_X\propto M^{4/3}(1+z)^{7/2}$ for $\Omega_m=1$ (Kaiser
1986, see Eke et al. 1998, for an extension to low--$\Omega_m$
cosmologies). As we also discussed in the introduction, this
prediction is at variance with respect to observational evidence of a
 steeper relation, $L_X\propto T^{\sim 3}$  for $T\magcir 2$ keV and,
possibly, even steeper for colder systems. This result is also in line
with the observed slope of the $L_X$--$M$ relation, $L_X\propto
M^\alpha$ with $\alpha\simeq 1.8\pm 0.1$ (Reiprich \& B\"ohringer
2002).

The first determinations of the \lt relation for clusters showed that
it has a quite large scatter (e.g. David et al. 1993, White, Jones \&
Forman 1997). A significant part of this has been recognized to be the
effect of cooling: central spikes associated with cooling regions
provide a large fraction of total $X$--ray luminosity, so that
differences in the cooling structure among clusters of similar
temperature induce a spread in the corresponding $L_X$ values. After
correcting for this effect, different authors (Allen \& Fabian 1998,
Markevitch 1998, Arnaud \& Evrard 1999) were able to calibrate a much
tighter \lt relation. Allen \& Fabian (1998) also noticed that the \lt
relation for the hottest clusters tends to flatten to the
self--similar scaling prediction, possibly suggesting that the effect
of extra heating becomes negligible for such systems.

\begin{figure} 
\centerline{ \hbox{\psfig{file=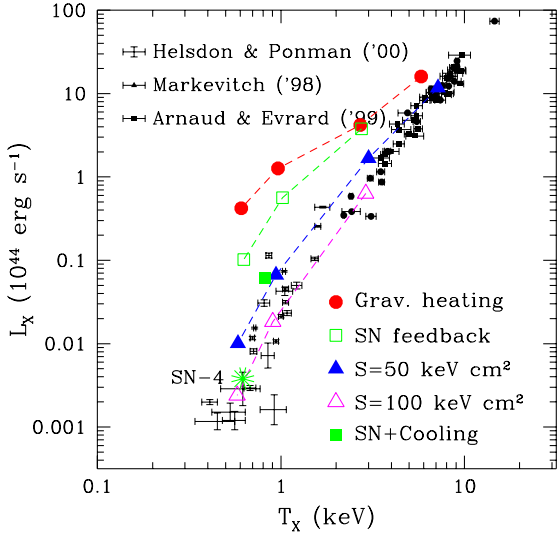,width=8cm} } }
\caption{The relation between $X$--ray luminosity and
emission--weighted temperature at $z=0$. Results from simulations are
compared to observational data for nearby clusters (data points with
errorbars) from different authors. The asterisk indicates the
SN-4 Hickson run, which corresponds to $E_h=1.4$ keV/part for the
heating energy.}
\label{fi:lt}
\end{figure}

Figure \ref{fi:lt} shows the effect of  extra heating on  the
\lt relation of nearby clusters. The GH runs are clearly at variance
with respect to data over the whole sampled range of temperatures.
Heating with the SN-1 recipe provides some steepening of the \lt
relation at the group scale, although not enough to reach agreement
with data points. A larger suppression of $L_X$ is achieved with the
S-50 and S-100 heating schemes and, for the Hickson group,
with the SN-4 run.  Quite interestingly, the SN-4 heating scheme
requires the same heating energy, $E_h\simeq 1.4$ keV/part as the S-50
entropy floor, but provides a significantly smaller luminosity, as a
consequence of the lower gas density in the central region of the
Hickson group (see also Fig. \ref{fi:fbar}). Therefore, a better
efficiency is reached in this case by gradually dumping energy within
the virialized regions of the ICM, rather than imposing an impulsive
pre--collapse heating on  the whole turn--around region.

\begin{figure} 
\centerline{ \hbox{\psfig{file=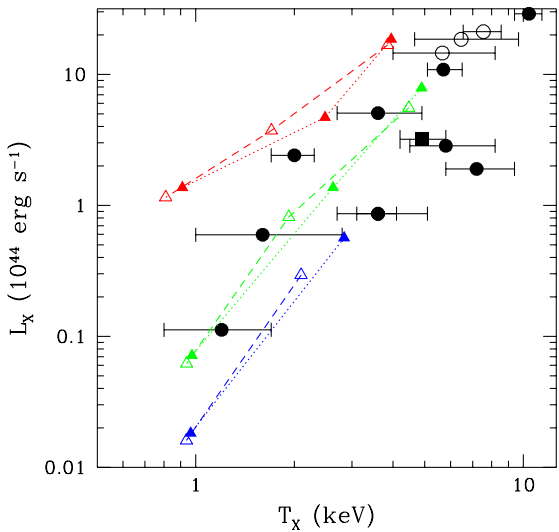,width=8cm} } }
\caption{The $L_X$--$T$ relation for distant $z\ge 0.5$,
clusters. Triangles connected by lines are simulation results for
Coma, Virgo and Fornax runs at $z=0.5$ (filled triangles with dotted
lines) and $z=1$ (open triangles with dashed lines). The three groups
of simulation results at decreasing $L_X$ refer to the GH, S-50 and
S-100 runs (for reasons of clarity we do not include the case of SN
feedback). Data points with errorbars refer to $z>0.5$ clusters from
Beppo--SAX and ASCA data (Della Ceca et al. 2000, open circles),
Chandra data extending out to $z=1.27$ (Borgani et al. 2001b, Stanford
et al. 2002, filled circles) and from the XMM observation of the
$z=1.26$ cluster of the Lockman hole (Hashimoto et al. 2002, square).}
\label{fi:lt_hz}
\end{figure}

In a similar way, Figure \ref{fi:lt_hz} shows a similar
comparison with observations for distant, $0.5\mincir z \mincir
1.3$,
clusters. Independent analyses have confirmed that data on \lt
relation of distant clusters are consistent with a lack of evolution
(e.g. Mushotzky \& Scharf 1997, Donahue et al. 1999, Della Ceca et
al. 2000, Borgani et al. 2001b, Stanford et al. 2002, Holden et
al. 2002). Since no data are available on $T\mincir 1$ keV groups in
the above $z$--range, we do not include in this comparison results for
the Hickson runs. Although temperature determinations for distant
clusters are prone to larger errorbars, the results are quite
in line with what shown in Fig. \ref{fi:lt}: the \lt relation of
distant clusters requires pre--heating of the ICM with $E_h\magcir 1$
keV/part.

\subsection{The effect of changing the epoch of heating}
Imposing an entropy floor at some pre--collapse redshift is a useful
approximation, in that it allows the characterization of the ICM
evolution by
a single quantity which is conserved in
adiabatic processes. As already mentioned, the heating energy required
to establish a given entropy floor is a non--trivial function of the
heating redshift, $z_h$, through the evolution of $\bar \rho$ and
$\delta_{\rm gas}$ (see eq.\ref{eq:entr1}). In general, one may also
expect
the resulting global ICM properties to depend on $z_h$. In
their simulations, Bialek et al. (2001) heated at a very high
redshift, $z_h\sim 20$, and found results on the required $S_{\rm fl}$
to reproduce the observed \lt relation which are generally consistent
with ours. This may suggest that the choice for $z_h$ is not
relevant. However, a close comparison between their runs and ours is
not straightforward: mass and dynamical resolutions are higher in our
runs and, as discussed in Section 2, this is likely to affect
observable quantities, such as $L_X$, which are  sensitive to the
details of gas clumpiness.

\begin{figure} 
\centerline{ \hbox{
\psfig{file=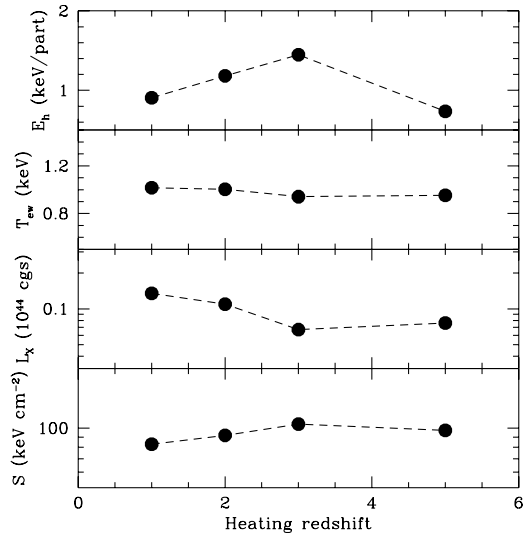,height=8cm,width=8cm,angle=0} } }
\caption{The effect of changing the heating redshift, $z_h$, for the
low--resolution S-50 Fornax run. Results are shown for the specific
heating energy, the emission--weighted temperature, the $X$--ray
luminosity and the entropy at $0.1\,R_{\rm vir}$ (from upper to lower
panels).}
\label{fi:zh} 
\end{figure}

As shown in Figure \ref{fi:zh}, the required $E_h$ for the S-50 Fornax
run changes by about a factor 2, $z_h=3$ being the less efficient
choice in terms of energy budget. Both $T_{ew}$ and central gas
entropy at $0.1\,R_{\rm vir}$ are quite stable, with variations of at
most $\sim 10\%$ and $\sim 20\%$ respectively. As expected, the
$X$--ray luminosity is the most sensitive quantity to changes of
$z_h$; it decreases by about a factor two as the heating epoch is
pushed back in time from $z_h=1$ to 5. Such relatively modest
variations suggest that different choices for $z_h$ induces
only moderate changes in the value of $S_{\rm fl}$ required to
reproduce the observed \lt relation.

\section{The effect of cooling and star formation}
\label{par:cool}
The simulations discussed so far have been performed  with the aim of
understanding in detail the effect of non--gravitational heating on
ICM observable quantities. However, excluding cooling in our
simulations is clearly a simplification, and its inclusion is likely
to change some results significantly. Cooling times can be
significantly shorter than the Hubble time in central regions of
clusters, thus causing a significant fraction of the gas there to cool
down and drop out of the hot diffuse, $X$--ray emitting phase (see
Fabian 1994, for a review on cooling processes in clusters).

\begin{figure} 
\centerline{ \hbox{
\psfig{file=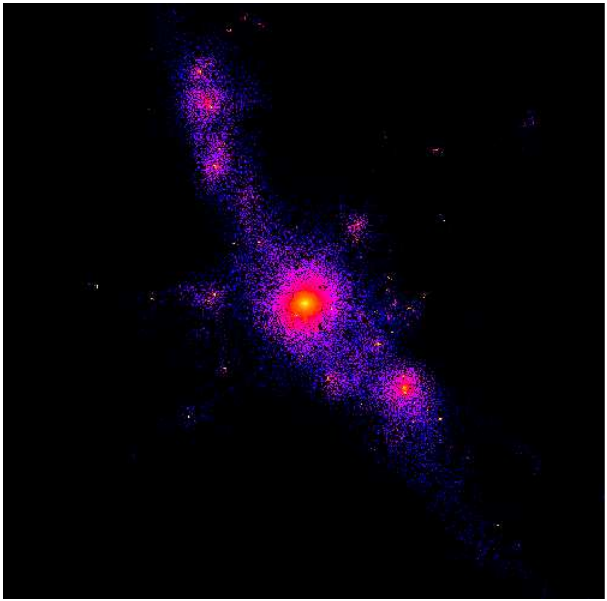,width=8.truecm} }}
\caption{The map of the gas density for the SN-1 Hickson run including
cooling and star formation. The box size is the same as in Fig. 2.} 
\label{fi:mapcool} 
\end{figure}

Cooling has been suggested as an alternative to non--gravitational
heating for breaking ICM self--similarity.  As gas undergoes cooling
in central cluster regions, $X$--ray luminosity is suppressed as a
consequence of the reduced amount of diffuse hot gas (Pearce et
al. 2000), thus possibly producing a steepening of the \lt relation
(Muanwong et al. 2001, Wu \& Xue 2002). The decreased pressure support
causes the recently shocked, higher entropy external gas to flow in,
thus causing a net increase of entropy for the hot gas left in central
cluster regions (Bryan 2000). Recently, Voit \& Bryan (2001) have
argued that only gas with entropy in excess of 100--200 keV cm$^2$
has
long enough cooling time to remain in the diffuse phase. Accordingly,
the entropy excess in central regions of poor groups should be
interpreted as an effect of cooling, rather than of extra heating.

\begin{figure}
\centerline{ \hbox{
\psfig{file=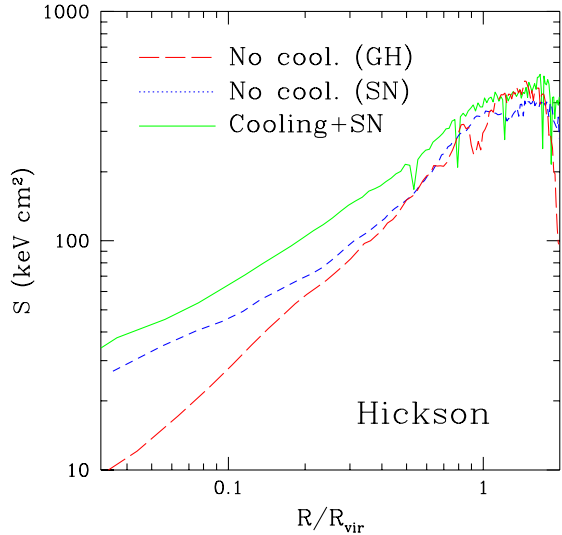,width=8cm} } }
\caption{The entropy profiles for the Hickson simulation without
cooling (for the GH and SN-1 cases) and including the effect of
cooling and star formation in the SN-1 run. In the computation of the
entropy for the cooling run, only gas particles belonging to the hot
diffuse phase have been considered (see text).}
\label{fi:profc} 
\end{figure}

In order to check this effect, we run  a high--resolution simulation of
the Hickson group, with the SN-1 heating scheme, but also including
gas cooling. The cooling function adopted assumes gas of primordial
composition, with zero metallicity, and follows the recipe of Wadsley
\& Bond (1997). We also include heating by UV background by adopting
the latest version of the model by Haardt \& Madau (2001). We follow
the recipe by Katz, Hernquist \& Weinberg (1992) to convert dense cold
gas particles into collisionless stars. Furthermore, when computing
properties of the diffuse ICM, we exclude gas particles with
temperatures $T<3\times 10^4$ K and overdensities $\delta>500$ (see
also Kay et al. 2002). These particles are assumed to belong to the
cold medium, although not yet removed from the gas phase by the star
formation (SF) algorithm. The effect of including cooling and SF is
that of collecting gas into dense knots, while suppressing gas density
in the diffuse filamentary structures (see Figure
\ref{fi:mapcool}). The resulting density profile in central regions
turns out to be steeper, $\propto R^{-\alpha}$, with $\alpha \simeq 2$
than that for a CDM only run (see also Treu \& Koopmans 2002).

In Figure \ref{fi:profc} we show the resulting entropy profile, as
compared to the GH and SN-1 runs, with no cooling. We detect some
increase of gas entropy inside $R_{vir}$, due to the infall of shocked
external gas. However, the inclusion of cooling and SF do not produce
any entropy plateau at $\sim 100$ keV cm$^2$. Although the entropy
increase goes in the right direction, it may be marginal in accounting
for the observed entropy excess in central group regions (e.g. Ponman
et al. 1999; see also Paper I).  A higher entropy could be possibly
attained by assuming a non--negligible metallicity in the cooling
function, which would cause an increasing efficiency of gas cooling.

Besides suppressing the amount of hot $X$--ray emitting gas, cooling
also causes the formation of high--density concentration of star
particles in the innermost cluster region. Such concentrations deepens
the potential well and causes temperature there to significantly
increase. Therefore, while the $X$--ray luminosity decreases, the
emission--weighted temperature increases. The effect of cooling in the
$M$--$T_{ew}$ and \lt relations of the Hickson run is shown with the
filled square in Figs. 9--11.  Such changes actually go in the right
direction of improving $L_X$--$T$ relation (see Figure \ref{fi:lt})
and possibly solve the discrepancy with the observed
$M_{500}$--$T_{ew}$ relation (see also Thomas et al. 2002; Voit et
al. 2002).

However, cooling in itself is a runaway process, due to the quick
increase of cooling efficiency with local gas density. As a
consequence, if not counteracted by some feedback process, a too large
fraction of gas can cool down to a collisionless phase (e.g., Prunet
\& Blanchard 1999, Lewis et al. 2000, Dav\'e et al. 2001, Balogh et
al. 2001). We see in our simulation that while the overall baryonic
fraction inside $R_{vir}$ at $z=0$ coincide with the cosmic value,
37\% of the gas is locked in the stellar phase.  This large fraction
is far in excess of the $\mincir 10\%$ fraction indicated by
observations (e.g., Balogh et al. 2001). Therefore, our SN feedback
scheme falls short in preventing the cooling runaway.
 
Addressing in detail the issue of preventing overcooling through a
suitable feedback mechanism is beyond the scope of this paper. Due to
the biased nature of halo formation in hierarchical models, the first
massive halos to collapse into a protocluster region will eventually
merge to form its core. These massive halos host the progenitors of
giant ellipticals in today clusters (Governato et al. 2001, Springel
et al. 2001) and will likely contribute to most of the energy released
in the ICM. Whatever the astrophysical source for this energy is, it
should act in a self--regulated way: cooling switches on the feedback
mechanism and the subsequent energy release inhibits too much gas to
leave the hot phase.  Besides SN, also AGN activity has been suggested
as a possible way of preventing overcooling, with energy released with
a duty cycle of a few $10^7$ years (e.g. Binney \& Tabor 1995, Ciotti
\& Ostriker 2001, Quilis et al. 2001, B\"ohringer et al. 2002).  If
AGN are responsible for the required feedback, then one should expect
to see some signature of their recent activity in the central regions
of those clusters whose cooling has been recently switched off.  A
succesfull and self--consistent inclusion of cooling and feedback in
numerical simulations has still to be realized, the main difficulty
lying in the treatment of physical processes which involve a large
dynamical range.  The physics describing AGN activity or SN explosions
involve small scales that are out of the reach of current and
foreseeable simulations of galaxy clusters, thus leaving as the only
possibility the inclusion of such process as external recipes.

\section{Conclusions}
We presented results from high resolution Tree+SPH simulations of
galaxy clusters and groups aimed at studying the effect of
non--gravitational heating of the intra--cluster medium (ICM) on
observable $X$--ray properties of galaxy systems. Four simulated
structures have been chosen so as to encompass the temperature range
from a poor Hickson--like group to a rich Coma--like cluster.  Two
different kinds of non--gravitational gas heating have been chosen:
{\em (i)} impulsive heating to set an entropy floor at
some redshift ($z_h=3$ as a reference epoch); {\em (ii)} gradual
heating, which follows the star--formation rate inside the cluster
region, as predicted by semi--analytical modelling of galaxy
formation, so as to mimic the effect of SN energy feedback.  Our main
results can be summarized as follows.
\begin{description}
\item[(a)] In order for simulations to reproduce analytical
predictions for the $L_X$--$T$ relation of gas sitting in hydrostatic
equilibrium within a NFW potential well, the structure of the gas
distribution needs to be resolved with at least 5--$10\times 10^4$ gas
particles within the virial radius, with a spatial resolution of the
order of 1--2\% of $R_{vir}$. 
\item[(b)] The observed slope and amplitude of the $L_{bol}$--$T_{ew}$
relation are well reproduced by setting an entropy floor of 50--100
keV cm$^2$. Assuming negligible radiative losses, SN feedback provides
$\simeq 0.35$ keV/part and, as such, has a too small effect on the \lt
relation. A four times larger energy, possibly consistent with that
obtainable from AGN, is required to reproduce the \lt relation at the
scale of a Hickson group.
\item[(c)] In the absence of cooling, the $M_{500}$--$T_{ew}$ relation
is found to be higher by $\simeq 40\%$ than that observed, over the
whole sampled $T$--range, almost independent of the presence and
amount of non--gravitational heating. Therefore, non--gravitational
heating in itself is not able to account at the same time for the
$L_{bol}$--$T_{ew}$ and the $M_{500}$--$T_{ew}$ relations. 
\item[(d)] The effect of including cooling in the simulation of a
Hickson group with SN heating is to suppress $L_X$ by about
40\% and increase $T_{ew}$ by about 30\%.  Therefore, the effect of
cooling goes in the right direction of decreasing the amplitude of the
\lt and $M_{500}$--$T_{ew}$ relations at the scale of groups. 
However, the runaway nature of cooling is not efficiently
counteracted by our SN feedback scheme, so that 37\% of the gas drops
out of the hot $X$--ray emitting phase. Such a large amount of cold
gas is in contrast with observational data on the stellar content of
galaxy systems. 
\end{description}

The results presented in this paper demonstrate that no unique simple
recipe exists to explain $X$--ray observable properties of galaxy
systems. Supernova heating provides a sizeable, but insufficient,
fraction of the non--gravitational heating energy required. Further
feedback energy from AGN may be required. Yet, the mechanisms to
thermalize the energy associated with QSO activity in the diffuse
medium has to be understood in detail. 

While much further work is needed, our study suggests that a
combination of different heating sources (SN and AGN) {\it and}
cooling will hopefully be able to reproduce {\it both} the correct
$L_X$--$T_{ew}$ and $M_{500}$--$T_{ew}$ relations, and to avoide gas
overcooling. A self--consistent treatment of such complex physics is
mandatory if the numerical description of the ICM has to keep pace
with the progress in the observational description of the hot baryons
in clusters.

\section*{Acknowledgements}
Simulations were run at the CINECA Supercomputing Center, with CPU
time provided by a grant of the Italian National Council for Astronomy
and Astrophysics (CNAA). We thank the referee, Jerry Ostriker, for
useful comments. We thank Fabrizio Brighenti and Mark Voit for reading
the paper and Sabrina De Grandi and Alexis Finoguenov for providing us
with the files of the data points shown in Figs. \ref{fi:tproj} and
\ref{fi:mtvir}, respectively. We acknowledge useful discussions with
Richard Bower, Stefano Ettori and Colin Norman. FG is supported in part
by the Brooks Fellowship.


\begin{thebibliography}{}
\bibitem[]{} Allen S.W., Fabian A.C., 1998, MNRAS, 297, 63
\bibitem[]{} Allen S.W., Schmidt R.W., Fabian A.C., 2001, MNRAS,
328, L37
\bibitem[]{} Anninos P., Norman M., 1996, 459, 12
\bibitem[]{} Arnaud M., Evrard A.E., 1999, MNRAS, 305, 631
\bibitem[]{} Arnaud M., Neumann D.M., Aghanim N., Gastaud R.,
Majerowicz S., Hughes J.P., 2001, A\&A, 365, L80
\bibitem[]{} Balogh M.L., Babul A., Patton D.R., 1999, MNRAS, 307, 463
\bibitem[]{} Balogh M.L., Pearce F.R., Bower R.G., Kay S.T., 2001,
MNRAS, 326, 1228
\bibitem[]{} Bialek J.J., Evrard A.E., Mohr J.J., 2001, ApJ, 555, 597 
\bibitem[]{} Bingelli B., Tammann G.A., Sandage A., 1987, AJ, 94, 251 
\bibitem[]{} Binney J., Tabor, G., 1995, MNRAS, 276, 663 
\bibitem[]{} B\"ohringer H., Matsushita K., Churazov E., Ikebe Y., Chen
Y., 2002, A\&A, 382, 804
\bibitem[]{} Borgani S., Governato F., Wadsley J., et
al., 2001a, ApJ, 559, L71 (Paper I)
\bibitem[]{} Borgani S., Rosati P., Tozzi P., et al., 2001b, ApJ, 561,
13
\bibitem[]{} Bower R.G., 1997, MNRAS, 288, 355
\bibitem[]{} Bower R.G., Benson A.J., Bough C.L., Cole S., Frenk C.S.,
Lacey C.G., 2001, MNRAS, 325, 497
\bibitem[]{} Boyle  B.J., Terlevich R.J., 1998, MNRAS, 293, L49
\bibitem[]{} Briel U.G., et al., 2001, A\&A, 365, L60
\bibitem[]{} Brighenti F., Mathews W.G., 2001, ApJ, 553, 103
\bibitem[]{} Bryan G.L., 2000, ApJ, 544, L1
\bibitem[]{} Bryan G.K., Norman M.L., 1998, ApJ, 495, 80
\bibitem[]{} Burles S., Tytler D., 1998, Space Sc. Rev., 84, 65 
\bibitem[]{} Cavaliere A., Menci N., Tozzi P., 1998, ApJ, 501, 493
\bibitem[]{} Cavaliere A., Menci N., Tozzi P., 1999, MNRAS, 308, 599 
\bibitem[]{} Cavaliere A., Padovani P., 1989, ApJ, 340, L5
\bibitem[]{} Ciotti, L., Ostriker, J.P., 2001, ApJ, 551, 131
\bibitem[]{} Cole S., Aragon-Salamanca A., Frenk C.S., Navarro J.F.,
Zepf, S.E., 1994, MNRAS, 271, 781
\bibitem[]{} Cole, S., Lacey, C., Baugh, C., Frenk, C., 2000, MNRAS,
319, 168
\bibitem[]{} Couchman, H.M.P. 1991, ApJL, 368, 23
\bibitem[]{} Dav\'e R., et al. 2001, ApJ, 552, 473
\bibitem[]{} David L.P., Slyz A., Jones C., Forman W<., Vrtilek S.D.,
Arnaud K.A., 1993, ApJ, 412, 479
\bibitem[]{} De Grandi S., Molendi S., 2002, ApJ, 567, 163
\bibitem[]{} Della Ceca R., Scaramella R., Gioia I.M., Rosati P.,
Fiore F., Squires G., 2000, A\&A, 353, 498
\bibitem[]{} Donahue M., Voit G.M., Scharf C.A., Gioia I.M.,
Mullis C.R., Hughes J.P., Stocke J.T., 1999, ApJ, 527, 525
\bibitem[]{} Drinkwater, M.J., et al. 2000, A\&A 355, 900
\bibitem[]{} Eke V.R., Cole S., Frenk C.S., 1996, MNRAS, 282, 263
\bibitem[]{} Eke V.R., Navarro J., Frenk C.S., 1998, ApJ, 503, 569
\bibitem[]{} Eke V.R., Navarro J., Steinmetz M., 2001, 554, 114
\bibitem[]{} Ettori S., 2002, MNRAS, 323, L1
\bibitem[]{} Ettori S., De Grandi S., Molendi S., 2002, A\&A, submitted
\bibitem[]{} Ettori S., Fabian A.C., 2001, in proceeding of the
Vulcano Workshop on ``Chemical Enrichment of the Intracluster and
Intergalactic Medium'',  eds. F. Matteucci, R. Fusco-Femiano, p. 23
\bibitem[]{} Evrard A.E., 1997, MNRAS, 292, 289
\bibitem[]{} Evrard A.E., Henry J.P., 1991, ApJ, 383, 95
\bibitem[]{} Evrard A.E., Metzler C.R., Navarro J.F., 1996, ApJ,
469, 494
\bibitem[]{} Fabian A.C., 1994, ARAA, 32, 277
\bibitem[]{} Fairley B.W., Jones L.R., Scharf C., Ebeling H., 
Perlman E., Horner D., Wegner G., Malkan M., 2000, MNRAS, 315, 669 
\bibitem[]{} Finoguenov A., Ponman T.J., 1999, MNRAS, 305, 325
\bibitem[]{} Finoguenov A., Reiprich T.H., B\"ohringer H., 2001, A\&A,
369, 479
\bibitem[]{} Finoguenov A., Jones C., B\"ohringer H., Ponman T.J.,
2002, ApJ, submitted
\bibitem[]{} Franceschini A., Hasinger G., Miyaji T., Malquori D.,
1999, MNRAS, 310, L5
\bibitem[]{} Frenk C.S., White S.D.M., Bode P., Bond J.R., Bryan G.L., et
al., 2000, ApJ, 525, 554
\bibitem[]{} Geller M.J., Diaferio A., Kurtz M.J., 1999, ApJ, 517, L23
\bibitem[]{} Ghigna S., Moore B., Governato F., Lake G., Quinn T.,
Stadel J., 2000, ApJ, 544, 616
\bibitem[]{} Governato F., Ghigna S., Moore B., Quinn T., Stadel J.,
Lake G., 2001, ApJ, 547, 555
\bibitem[]{} Haardt F., Madau P., 2001, in Proceedings of the XXXVIth
Rencontres de Moriond, eds. D.M. Neumann et al. (preprint
astro--ph/0106018)
\bibitem[]{} Hashimoto Y., Hasinger G., Aranud M., Rosati P., Miyaji
T., 2002, A\&A, 381, 841
\bibitem[]{} Helsdon S.F., Ponman T.J., 2000, MNRAS, 315, 356 
\bibitem[]{} Holden B., Stanford S.A., Rosati P., Squires G.K., Tozzi
P., Eisenhardt P., Elston R., 2002, ApJ, in press (preprint
astro--ph/0203474) 
\bibitem[]{} Horner D.J., Mushotzky, R.F., Scharf C.A., 1999, ApJ,
520, 78 
\bibitem[]{} Kaiser N., 1986, MNRAS, 222, 323
\bibitem[]{} Kaiser N., 1991, ApJ, 383, 104
\bibitem[]{} Katz, N., Hernquist, L., Weinberg, D.H., 1992, ApJS, 105,
19 
\bibitem[]{} Katz N., White S.D.M., 1993, ApJ, 412, 455
\bibitem[]{} Kauffmann G., White S.D.M., Guiderdoni B., 1993, MNRAS, 
264, 201
\bibitem[]{} Kay S.T., Pearce F.R., Frenk C.S., Jenkins A., 2002,
MNRAS, 330, 113
\bibitem[]{} Kitayama T., Suto J., 1996, ApJ, 469, 480
\bibitem[]{} Kravtsov A.V., Yepes G., 2000, MNRAS, 318, 227
\bibitem[]{} Ikebe Y., Reiprich T.H., B\"ohringer H., Tanaka, Y.,
Kitayama T., 2002, A\&A, 383, 773
\bibitem[]{} Irwin J.A., Bregman J.N., 2000, ApJ, 538, 543
\bibitem[]{} Lacey C., Cole S., 1993, MNRAS, 262, 627
\bibitem[]{} Lewis G.F., Babul A., Katz N. Quinn T., Hernquist L.,
Weinberg D.H., 2000, ApJ, 536, 623
\bibitem[]{} Lloyd-Davies E.J., Ponman T.J., Cannon D.B., 2000, MNRAS,
315, 689
\bibitem[]{} Loewenstein M., 2001, ApJ, 557, 573
\bibitem[]{} Loewenstein M., Mushotzky R., 1996, ApJ, 466, 695
\bibitem[]{} Markevitch M., 1998, ApJ, 504, 27
\bibitem[]{} Markevith M., Forman W.R., Sarazin C.L., Vikhlinin A.,
1998, ApJ, 503, 77
\bibitem[]{} Markevitch M., et al., 2000, ApJ, 541, 542
\bibitem[]{} Mazzotta P., Markevitch M., Vikhlinin A., Forman W.R.,
David L.P., VanSpeybroek L., 2001, ApJ, 555, L87
\bibitem[]{} McNamara B.R., Wise M., Nulsen P.E.J., David L.P.,
Sarazin C.L., Bautz M., Markevitch M., Vikhlinin A., Forman W.R.,
Jones C. \& Harris D.E., 2000 ApJ, 534, L135
\bibitem[]{} Menci N., Cavaliere A., 2000, MNRAS, 311, 50
\bibitem[]{} Mo H.J., Mao S., White S.D.M., 1998, MNRAS, 295, 319  
\bibitem[]{} Moore B., Governato F., Quinn T., Stadel J., Lake G.,
1998, ApJ, 499, L5
\bibitem[]{} Muanwong O., Thomas P.A., Kay S.T., Pearce F.R., Couchman
H.M.P., 2001, ApJ, 552, L27
\bibitem[]{} Mushotzky R.F., Scharf C.A., 1997, ApJ, 482, L13
\bibitem[]{} Nath B.B., Roychowdhury S., 2002, MNRAS, 333, 145
\bibitem[]{} Navarro J.F., Frenk C.S., White S.D.M., 1995, MNRAS, 275,
720 
\bibitem[]{} Navarro J.F., Frenk C.S., White S.D.M., 1997, ApJ, 490, 493
\bibitem[]{} Neumann D.M., Arnaud M., 2001, A\&A, 365, L80
\bibitem[]{} Nevalainen J., Markevitch, M., Forman W.R., 2000, ApJ,
536, 73
\bibitem[]{} Pearce F.R., Thomas P.A., Couchman H.M.P., Edge A.C.,
2000, MNRAS, 317, 1029
\bibitem[]{} Pierpaoli E., Scott D., White M., 2001, MNRAS,
325, 77
\bibitem[]{} Pildis R.A., Bregman J.N., Evrard A.E., 1995, ApJ, 443,
514 
\bibitem[]{} Pipino A., Matteucci F., Borgani S., Biviano A., 2002,
NewA, in press (preprint astro--ph/0204161)
\bibitem[]{} Poli, F., Giallongo, E., Menci, N., D'Odorico, S., \& Fontana,
A. 1999, ApJ, 527, 662
\bibitem[]{} Ponman T.J., Bourner P.D.J., Ebeling H., B\"ohringer H.,
1996, 293, 690
\bibitem[]{} Ponman T.J., Cannon D.B., Navarro J.F., 1999, Nature,
397, 135
\bibitem[]{} Power C., Navarro J.F., Jenkins A., Frenk C.S., White
S.D.M., Springel V., Stadel J., Quinn T., 2002, MNRAS, submitted
(preprint astro--ph/0201544)
\bibitem[]{} Prunet S., Blanchard A., 1999, A\&A, submitted (preprint
astro--ph/9909145) 
\bibitem[]{} Quilis V., Bower R.G., Balogh M.L., 2001, MNRAS, 328,
1091 
\bibitem[]{} Raymond J.C., Smith B.W., 1977, ApJS, 35, 419
\bibitem[]{} Reiprich T.H., B\"ohringer H., 2002, ApJ, 567, 716
\bibitem[]{} Renzini A., 1997, ApJ, 488, 35
\bibitem[]{} Salpeter E.E., 1955, ApJ, 121, 161
\bibitem[]{} Seljak U., 2002, MNRAS, submitted (preprint
astro--ph/0111362)
\bibitem[]{} Somerville R.S., Primack J.R., 1999, MNRAS, 310, 1087
\bibitem[]{} Springel V., White S.D.M., Tormen G., Kauffmann G., 2001,
MNRAS, 328, 726
\bibitem[]{} Stadel, J., 2001, Ph. D. thesis.
\bibitem[]{} Stanford S.A., Holden B.P., Rosati P., Eisenhardt P.R.,
Stern D., Squires G., Spinrad H., 2002, ApJ, in press (preprint
astro--ph/0110709)
\bibitem[]{} Steinmetz M., White S.D.M., 1997, MNRAS, 288, 545
\bibitem[]{} Suginohara, T., Ostriker, J.P., 1998, ApJ, 507, 16
\bibitem[]{} Thomas P.A., Muanwong O., Kay, S.T., Liddle, A.R., 2002,
MNRAS, 330, L48
\bibitem[]{} Tozzi P., Norman C. 2001, ApJ, 546, 63 (TN01)
\bibitem[]{} Treu, T., Koopmans, L., 2002, ApJ, submitted (preprint
astro--ph/0202342) 
\bibitem[]{} Valageas P., Silk J., 1999, A\&A, 350, 725
\bibitem[]{} Valdarnini R., 2002, ApJ, 567, 741
\bibitem[]{} Viana P.T.P., Nichol R.C., Liddle A.R., 2002, ApJ,
569, L75
\bibitem[]{} Voit G.M., Bryan G.L., 2001, Nature, 414, 425
\bibitem[]{} Voit G.M., Donahue M., 1998, ApJ, 500, L111
\bibitem[]{} Voit, G.M., Bryan, G.L., Balogh, M.L., Bower, R.G., 2002,
ApJ, submitted
\bibitem[]{} Wadsley, J.W., Bond, J.R., 1997, Computational
Astrophysics; 12th Kingston Meeting on Theoretical Astrophysics,
p. 332 (astro--ph/9612148)
\bibitem[]{} White D.A., 2000, MNRAS, 312, 663
\bibitem[]{} White D.A., Jones C., Forman W., 1997, MNRAS, 292, 419
\bibitem[]{} White S.D.M., Efstathiou G., Frenk C.S., 1993, MNRAS,
262, 1023
\bibitem[]{} White S.D.M., Frenk C.S., 1991, ApJ, 379, 52
\bibitem[]{} Wu K.K.S., Fabian A.C., Nulsen P.E.J., 1998, MNRAS, 301, L20 
\bibitem[]{} Wu K.K.S., Fabian A.C., Nulsen P.E.J., 2000, MNRAS,
318, 889
\bibitem[]{} Wu X.-P., Xue Y.-J., 2002, ApJ, 569, 112
\bibitem[]{} Yamada M., Fujita Y., 2001, ApJ, 553, 145
\bibitem[]{} Yoshida N., Springel V., White S.D.M., Tormen G., 2001,
535, L103
\end{thebibliography}
\end{document}